\documentclass[12pt,a4paper]{article}
\pdfoutput=1
\usepackage{amsmath,amssymb,bbm,slashed,cancel}
\usepackage{graphicx}
\usepackage{slashed}
\usepackage{hyperref}
\usepackage{subcaption}
\usepackage{cite}
\usepackage{dsfont}
\usepackage{stackengine}
\usepackage{accents}
\usepackage[utf8]{inputenc}

\usepackage{colortbl,colordvi,xcolor,color}

\usepackage{array,multirow}
\usepackage{booktabs}
\usepackage{float}

\allowdisplaybreaks

\newcommand{\eg}{{\em e.g.}}
\newcommand{\ie}{{\em i.e.}}

\newcommand\unit[1]{\,\mathrm{#1}}
\newcommand\GeV{\unit{GeV}}
\newcommand\TeV{\unit{TeV}}
\newcommand\olrarrow[1]{\overset{\text{\scriptsize$\leftrightarrow$}}{#1}}

\newcommand{\crowcolor}{\rowcolor[rgb]{0.9,0.9,0.9}}

\makeatletter
\newcommand\xleftrightarrow[2][]{%
  \ext@arrow 9999{\longleftrightarrowfill@}{#1}{#2}}
\newcommand\longleftrightarrowfill@{%
  \arrowfill@\leftarrow\relbar\rightarrow}
\makeatother

\begin{document}

\begin{titlepage}
\setcounter{page}{0} 
\begin{flushright}
KEK--TH--2215
\end{flushright}
\vskip 1.5cm
\begin{center}
  {\Large \bf Muon $\boldsymbol{g-2}$ and CKM Unitarity \\\vspace{5mm} 
  in Extra Lepton Models
  }
\vskip 1.5cm
{
  Motoi Endo,$^{(a,b)}$
  and
  Satoshi Mishima$^{(a)}$
}

\vspace{1.5em}

\begingroup\small\itshape
$^{(a)}$ \!\! KEK Theory Center, IPNS, KEK, Tsukuba, Ibaraki 305--0801, Japan
\\[0.3em]
$^{(b)}$ \!\! The Graduate University of Advanced Studies (Sokendai), Tsukuba, Ibaraki 305--0801, Japan
\endgroup

\vspace{1cm}

\abstract{
\noindent 
We investigate the impact of extra leptons on observed tensions 
in the muon $g-2$ and the first-row CKM unitarity. 
By introducing a new SU(2)$_L$ doublet lepton and a SU(2)$_L$ triplet lepton, 
we find that both of the tensions can be explained simultaneously under constraints from electroweak precision observables and Higgs-boson decays. 
Our model could be tested by measurements of $h\to\mu\mu$ at future collider experiments. 
}

\end{center}
\end{titlepage}

\renewcommand{\thefootnote}{\#\arabic{footnote}}
\setcounter{footnote}{0}

\section{Introduction}

Flavor physics provides powerful probes for new physics (NP) beyond the Standard Model (SM). At present, some of flavor measurements show tensions with their SM predictions. In this paper, we investigate a tension in the anomalous magnetic moment of muon $a_\mu=(g_\mu-2)/2$, so-called the muon $g-2$, and that in the first-row unitarity of the Cabibbo-Kobayashi-Maskawa (CKM) matrix. They hint at NP that couples to muon.

The muon $g-2$ exhibits a long-standing difference between the experimental measurement and the theory prediction in the SM. The latest value of the SM prediction is~\cite{Aoyama:2020ynm}\footnote{ A recent lattice study on the leading-order hadronic vacuum-polarization contribution shows no tension in the muon $g-2$~\cite{Borsanyi:2020mff}. 
The result is inconsistent by $>3\sigma$ with those based on dispersive analyses for $e^+e^-\to$ hadron data~\cite{Davier:2019can,Keshavarzi:2019abf}.}
\begin{align}
\Delta a_\mu 
=
a_\mu^{\mathrm{exp}} - a_\mu^{\mathrm{SM}}
=
(27.9 \pm 7.6)\times 10^{-10},
\label{eq:deltaamu}
\end{align}
which corresponds to $3.7\,\sigma$ discrepancy. Here the experimental value is taken to be 
$a_\mu^{\mathrm{exp}}=(11\,659\,208.9 \pm 5.4 \pm 3.3)\times 10^{-10}$, which is calculated from the result of the E821 experiment~\cite{Bennett:2002jb,Bennett:2004pv,Bennett:2006fi} 
with the latest value of the muon-to-proton magnetic ratio in the CODATA 2018~\cite{CODATA2018}. 
This discrepancy implies the potential existence of NP coupled to muon.\footnote{The electron $g-2$ with a precision measurement of the fine structure constant using caesium atoms also shows a discrepancy: 
$\Delta a_e=(-0.88\pm 0.36)\times 10^{-12}$~\cite{Parker:2018vye}. We do not consider it in the current study.}

The recent studies on the CKM matrix elements, $V_{ud}$ and $V_{us}$, also show a tension with the CKM unitariry. 
The most precise determination of $|V_{ud}|$ comes at present from the superallowed $0^+\to 0^+$ nuclear $\beta$ decays~\cite{Hardy:2014qxa,Hardy:2016vhg,Hardy:2018zsb}. The extraction, however, suffers from theoretical uncertainty in the transition-independent part of hadronic contributions to electroweak (EW) radiative corrections~\cite{Marciano:2005ec}. 
Recent studies of them lead to 
\begin{align}
|V_{ud}|
= 
\begin{cases}
0.97370 \pm 0.00014 & \text{(SGPR)~\cite{Seng:2018yzq}}, \\
0.97389 \pm 0.00018 & \text{(CMS)~\cite{Czarnecki:2019mwq}},\\
0.97365 \pm 0.00015 & \text{(SFGJ)~\cite{Seng:2020wjq}},
\end{cases}
\label{eq:VudData}
\end{align}
which are consistent with each other. 
On the other hand, 
$|V_{us}/V_{ud}|$ and 
$|V_{us}|$ are extracted from 
the leptonic-decay ratio $K_{\mu 2}/\pi_{\mu 2}$ and 
the semileptonic decays $K_{\ell 3}$ ($\ell=e,\mu$), 
respectively~\cite{Cirigliano:2019,Passemar:2019}:
\begin{align}
\left|\frac{V_{us}}{V_{ud}}\right| = 0.23129\pm 0.00045,
\quad
|V_{us}| = 0.22326\pm 0.00058.
\end{align}
The measured values of $|V_{ud}|$, $|V_{us}/V_{ud}|$ and $|V_{us}|$ violate the first-row CKM unitariry~\cite{Belfatto:2019swo,Grossman:2019bzp}. Defining the amount of the violation as $|V_{ud}|^2+|V_{us}|^2+|V_{ub}|^2=1+\Delta_{\mathrm{CKM}}$ with  
$|V_{ub}| \approx 0.003683$~\cite{Charles:2004jd,CKMfitter:2019summer}, 
we have 
\begin{align}
  \Delta_{\mathrm{CKM}}
  &= 
  \begin{cases}
  -0.00118\pm 0.00034,
  & (\text{SGPR},~K_{\mu 2}/\pi_{\mu 2}), \\
  -0.00205\pm 0.00038,
  & (\text{SGPR},~K_{\ell 3}), \\
  -0.00079\pm 0.00040,
  & (\text{CMS},~K_{\mu 2}/\pi_{\mu 2}), \\
  -0.00168\pm 0.00044,
  & (\text{CMS},~K_{\ell 3}), \\
  -0.00128\pm 0.00036,
  & (\text{SFGJ},~K_{\mu 2}/\pi_{\mu 2}), \\
  -0.00215\pm 0.00039,
  & (\text{SFGJ},~K_{\ell 3}), \\
  \end{cases}
  \label{eq:CKMunitarity}
\end{align}
which are away from zero at the $3.5\,\sigma$, $5.4\,\sigma$,
$2.0\,\sigma$, $3.8\,\sigma$, $3.6\,\sigma$ and $5.5\,\sigma$ level, respectively.\footnote{
It is also noticed that the value of $|V_{us}|$ calculated by combining $|V_{us}/V_{ud}|$ from $K_{\mu 2}/\pi_{\mu 2}$ with $|V_{ud}|$ from the nuclear $\beta$ decays is in tension with that from $K_{\ell 3}$~\cite{Cirigliano:2019}. 
}
This violation may suggest a NP contribution to the $W$-$\mu$-$\nu$ interaction~\cite{Belfatto:2019swo,Coutinho:2019aiy,Crivellin:2020lzu}.

Both of the above tensions imply NP that couples to muon. 
The effective field theory analysis tells us its energy scale. 
The effective Lagrangian for the muon $g-2$, $\mathcal{L}_{\mathrm{eff}}= (1/\Lambda^2) (\bar\ell \sigma^{\mu\nu}\mu_{R}) \phi A_{\mu\nu} + \mathrm{h.c.}$, where $\phi$ is the SM Higgs doublet, implies the NP scale $\Lambda\sim 300$\,TeV to accommodate the tension in Eq.~\eqref{eq:deltaamu}. 
On the other hand, the NP contributions to the $W$-$\mu$-$\nu$ interaction is described by $\mathcal{L}_{\mathrm{eff}}= (1/\Lambda^2)(\phi^\dagger i \olrarrow{D_\mu^a} \phi) (\bar\ell \gamma^\mu \sigma^a \ell)$, where $\sigma^a$ are the Pauli matrices. 
The CKM tension in Eq.~\eqref{eq:CKMunitarity} implies $\Lambda\lesssim 10$\,TeV, which is one order of magnitude lower than the scale for the muon $g-2$. 
Namely, we expect that NP contributions to the CKM measurements are much larger than those to the muon $g-2$.

In this paper, we study extra lepton models as a candidate to solve the scale hierarchy in the NP contributions to the muon $g-2$ and the CKM measurements. 
The extra leptons can contribute to the latter at the tree level~\cite{delAguila:2008pw,Dermisek:2013gta}, while effects on the former arise first at the one-loop level~\cite{Czarnecki:2001pv,Kannike:2011ng,Dermisek:2013gta,Freitas:2014pua,Megias:2017dzd,Poh:2017tfo,Kowalska:2017iqv,Raby:2017igl,Calibbi:2018rzv,Crivellin:2018qmi,Kawamura:2019rth,Kawamura:2019hxp}.  
This explains naturally the hierarchy in the NP contributions. 
We investigate correlations between them
under constraints from EW precision observables (EWPO) and the Higgs boson decay into a muon pair.\footnote{
Constraints from the lepton-flavor-universarity violating ratios studied in Refs.~\cite{Coutinho:2019aiy,Crivellin:2020lzu} are weaker, and not considered in this study. 
}

This paper is organized as follows. In Section~\ref{sec:model} we present our extra lepton model and its matching to the SM effective field theory (SMEFT). 
In Sections~\ref{sec:EWPO} and \ref{sec:Higgs} 
we explain constraints from the EWPO and the Higgs boson decay, respectively. 
In Sections~\ref{sec:CKM} and \ref{sec:gmin2} we discuss extra lepton contributions to the CKM measurements and the muon $g-2$, respectively. 
In Section~\ref{sec:result} we present our numerical analysis. 
Finally our conclusions are drawn in Section~\ref{sec:conclusion}.

\section{Extra lepton model}
\label{sec:model}

\begin{table}[t]
\centering
\begin{tabular}{ccc c cccc}
\hline
\crowcolor
$\ell$ & $\mu_R$ & $\phi$ & \phantom{0} &
$E_{L,R}$ & $(\Delta_1)_{L,R}$ & $(\Delta_3)_{L,R}$ & $(\Sigma_1)_{L,R}$ \\
$2_{-\frac{1}{2}}$ &
$1_{-1}$ &
$2_{\frac{1}{2}}$ &
&
$1_{-1}$ &
$2_{-\frac{1}{2}}$ &
$2_{-\frac{3}{2}}$ &
$3_{-1}$ \\ [4pt]
\hline
\end{tabular}
\caption{List of particles in the model. The quantum numbers represent $(\mathrm{SU}(2)_L)_{\mathrm{U}(1)_Y}$.}
\label{tab:particle}
\end{table}

We introduce extra leptons which couple to the muons and have vectorlike masses.\footnote{
If the extra leptons couple the electron or tau leptons simultaneously, lepton flavor violations are induced. } 
The particle contents are summarized in Table \ref{tab:particle}.\footnote{
In addition, a gauge singlet $N \sim 1_{0}$ and an SU(2)$_L$ adjoint lepton $\Sigma \sim 3_{0}$ are not included in the table because they are likely to generate too large neutrino masses by the seesaw mechanisms~\cite{Minkowski:1977sc,GellMann:1980vs,Yanagida:1979as,Mohapatra:1979ia,Foot:1988aq}. }
Here, $\ell = (\nu_L,\mu_L)^T$ is the SM SU(2)$_L$ doublet lepton in the second generation, and $\mu_R$ is the right-handed muon singlet.
The Higgs doublet $\phi$ obtains a vacuum expectation value 
after the electroweak symmetry breaking (EWSB) as $\phi = \left[0,(v+h)/\sqrt{2}\right]^T$, where the Nambu–Goldston bosons are ignored.
The $\mathrm{SU}(2)_L$ multiplets of the extra leptons are explicitly shown as
\begin{align}
 & \Delta_1 = (\Delta_1^0,\Delta_1^-)^T,~~~
   \Delta_3 = (\Delta_3^-,\Delta_3^{--})^T, \\
 & \Sigma_1 = (\Sigma_1^1,\Sigma_1^2,\Sigma_1^3)^T
 = \left[
 \frac{  \Sigma_1^0 + \Sigma_1^{--} }{\sqrt{2}},
 \frac{i(\Sigma_1^0 - \Sigma_1^{--})}{\sqrt{2}},
 \Sigma_1^-\right]^T.
 \label{eq:sigma1}
\end{align}
In each field, the superscript $0,-,--$ denotes the electric charge $Q$. 
Besides, $Q=0$ for $\nu_L$ and $-1$ for $\mu_{L,R}$ and $E_{L,R}$.
The gauge interactions are represented as
\begin{align}
 \mathcal{L}_{\rm int} &= 
 e Q \bar f \gamma^\mu f A_\mu
 \notag \\ &\quad
 + \frac{g}{c_W} \bar f \gamma^\mu 
 \big[(T_L^{\prime 3} - Q s_W^2) P_L + (T_R^{\prime 3} - Q s_W^2) P_R \big] 
 f Z_\mu
 \notag \\ &\quad
 + \frac{g}{\sqrt{2}} ( 
 \bar \nu \gamma^\mu \mu_L + 
 \bar \Delta_{i}^0 \gamma^\mu \Delta_{i}^- +
 \sqrt{2}\, \bar \Sigma_{j}^0 \gamma^\mu \Sigma_{j}^- +
 \sqrt{2}\, \bar \Sigma_{j}^- \gamma^\mu \Sigma_{j}^{--}
 ) W_\mu^+ + {\rm h.c.},
\end{align}
where $A_\mu$, $Z_\mu$ and $W_\mu$ are the gauge bosons, and $f$ represents a fermion in Table \ref{tab:particle} with $i = 1L,1R,3L,3R$ and $j = 1L,1R$ in the last line.
Here and hereafter, $s_W = \sin\theta_W$ and $c_W = \cos\theta_W$ with the Weinberg angle $\theta_W$.
The SU(2)$_L$ charge $T_{L,R}^{\prime 3}$ is shown as\footnote{
It is noticed that the representation of $\Sigma_1$ in Eq.~\eqref{eq:sigma1} is not an eigenstate of the SU(2)$_L$ generator $\hat T^3$.
This is introduced to represent the Yukawa interactions in Eq.~\eqref{eq:Yukawa}. }
\begin{align}
 T_L^{\prime 3} = 
 \begin{cases}
 1 & \mathrm{for}~~\Sigma_{1L}^0, \\
 1/2 & \mathrm{for}~~\nu_L, \Delta_{1L}^0, \Delta_{3L}^-, \\
 0 & \mathrm{for}~~E_L, \Sigma_{1L}^-, \\
 -1/2 & \mathrm{for}~~\mu_L, \Delta_{1L}^-, \Delta_{3L}^{--}, \\
 -1 & \mathrm{for}~~\Sigma_{1L}^{--}, \\
 \end{cases}~~~~~
 T_R^{\prime 3} = 
 \begin{cases}
 1 & \mathrm{for}~~\Sigma_{1R}^0, \\
 1/2 & \mathrm{for}~~\Delta_{1R}^0, \Delta_{3R}^-, \\
 0 & \mathrm{for}~~\mu_R, E_R, \Sigma_{1R}^-, \\
 -1/2 & \mathrm{for}~~\Delta_{1R}^-, \Delta_{3R}^{--}, \\
 -1 & \mathrm{for}~~\Sigma_{1R}^{--}. \\
 \end{cases}
\end{align}

In general, the Yukawa interactions and vectorlike mass terms are given by
\begin{align}
 -\mathcal{L}_{\rm int} &= 
 y_\mu\, \bar \ell \phi \mu_R
 \notag \\ &~~~
 + \lambda_E\, \bar E_R \phi^\dagger \ell 
 + \lambda_{\Delta_1}\, \bar \Delta_{1L} \phi \mu_R 
 + \lambda_{\Delta_3}\, \bar \Delta_{3L} \tilde \phi \mu_R 
 + \lambda_{\Sigma_1}\, \bar \Sigma_{1R}^a \phi^\dagger \sigma^a \ell
 \notag \\ &~~~
 + \lambda_{E\Delta_1}\, \bar E_L \phi^\dagger \Delta_{1R}
 + \lambda_{\Delta_1E}\, \bar \Delta_{1L} \phi E_R
 \notag \\ &~~~
 + \lambda_{E\Delta_3}\, \bar E_L \tilde \phi^\dagger \Delta_{3R}
 + \lambda_{\Delta_3E}\, \bar \Delta_{3L} \tilde \phi E_R
 \notag \\ &~~~
 + \lambda_{\Sigma_1\Delta_1}\, \bar \Sigma_{1L}^a \phi^\dagger \sigma^a \Delta_{1R}
 + \lambda_{\Delta_1\Sigma_1}\, \bar \Delta_{1L} \sigma^a \phi \Sigma_{1R}^a
 \notag \\ &~~~
 + \lambda_{\Sigma_1\Delta_3}\, \bar \Sigma_{1L}^a \tilde \phi^\dagger \sigma^a \Delta_{3R}
 + \lambda_{\Delta_3\Sigma_1}\, \bar \Delta_{3L} \sigma^a \tilde \phi \Sigma_{1R}^a
 \notag \\ &~~~
 + M_E\, \bar E_L E_R
 + M_{\Delta_1}\, \bar \Delta_{1L} \Delta_{1R}
 + M_{\Delta_3}\, \bar \Delta_{3L} \Delta_{3R}
 + M_{\Sigma_1}\, \bar \Sigma_{1L}^a \Sigma_{1R}^a
 + {\rm h.c.},
 \label{eq:Yukawa}
\end{align}
where $\sigma^a$ are the Pauli matrices and $\tilde \phi = i\sigma^2 \phi^*$.
Here and hereafter, all the coupling constants are supposed to be real.
Besides, the Yukawa couplings $\lambda_E$, $\lambda_{\Delta_1}$, $\lambda_{\Delta_3}$, and $\lambda_{\Sigma_1}$ as well as the vectorlike masses $M_i$ are chosen to be positive by rotating fields without loss of generality.

After the EWSB, the mass term of the singly-charged leptons is obtained as
\begin{align}
 -\mathcal{L}_{m} &= 
 \begin{bmatrix}
 \bar \mu_L & \bar E_L & \bar \Delta_{1L}^- & \bar \Delta_{3L}^- & \bar \Sigma_{1L}^-
 \end{bmatrix}
 \mathcal{M}_-
 \begin{bmatrix}
 \mu_R \\ E_R \\ \Delta_{1R}^- \\ \Delta_{3R}^- \\ \Sigma_{1R}^-
 \end{bmatrix}
 +{\rm h.c.},
 \end{align}
 where the mass matrix $\mathcal{M}_-$ is given in terms of the Yukawa matrix $Y_-$ as  
 \begin{align}
 \mathcal{M}_- &=
 \frac{v}{\sqrt{2}}\, Y_-
 +
 \mathrm{diag}\Big(
 0,M_E, M_{\Delta_1}, M_{\Delta_3}, M_{\Sigma_1}
 \Big),
 \\[1mm]
 Y_- &=
 \begin{bmatrix}
 y_\mu & 
 \lambda_E &
 0 &
 0 &
 -\lambda_{\Sigma_1} \\
 0 & 
 0 &
 \lambda_{E\Delta_1} &
 \lambda_{E\Delta_3} &
 0 \\
 \lambda_{\Delta_1} & 
 \lambda_{\Delta_1E} &
 0 &
 0 &
 -\lambda_{\Delta_1\Sigma_1} \\
 \lambda_{\Delta_3} & 
 \lambda_{\Delta_3E} &
 0 &
 0 &
 \lambda_{\Delta_3\Sigma_1} \\
 0 & 
 0 &
 -\lambda_{\Sigma_1\Delta_1} &
 \lambda_{\Sigma_1\Delta_3} &
 0 \\
 \end{bmatrix}.
\end{align}
The matrix $\mathcal{M}_-$ is diagonalized by a biunitary transformation: 
\begin{align}
 U_L^{- \dagger} \mathcal{M}_- U_R^{-} = {\rm diag}(m_{i^-}).
\end{align}
Similarly, the mass terms of the neutral and doubly-charged leptons are written as 
\begin{align}
 -\mathcal{L}_{m} &= 
 \begin{bmatrix}
 \bar \nu_L & \bar \Delta_{1L}^0 & \bar \Sigma_{1L}^0
 \end{bmatrix}
 \mathcal{M}_0
 \begin{bmatrix}
 \nu_L^c \\ \Delta_{1R}^0 \\ \Sigma_{1R}^0
 \end{bmatrix}
  +
 \begin{bmatrix}
 \Delta_{3L}^{--} & \bar \Sigma_{1L}^{--}
 \end{bmatrix}
 \mathcal{M}_{--}
 \begin{bmatrix}
 \Delta_{3R}^{--} \\ \Sigma_{1R}^{--}
 \end{bmatrix}
 +{\rm h.c.}, 
 \\ 
 \mathcal{M}_0 &=
 \begin{bmatrix}
 0 & 0 & v\lambda_{\Sigma_1} \\
 0 & M_{\Delta_1} & v\lambda_{\Delta_1\Sigma_1} \\
 0 & v\lambda_{\Sigma_1\Delta_1} & M_{\Sigma_1}
 \end{bmatrix}, \quad
 \mathcal{M}_{--} =
 \begin{bmatrix}
 M_{\Delta_3} & v \lambda_{\Delta_3\Sigma_1} \\
 v \lambda_{\Sigma_1\Delta_3} & M_{\Sigma_1}
 \end{bmatrix}.
\end{align}
Note that the mass matrix $\mathcal{M}_0$ does not contribute to neutrino masses, \ie, avoiding too heavy neutrinos. 
These mass matrices are diagonalized as 
\begin{align}
 U_L^{0 \dagger} \mathcal{M}_0 U_R^{0} = {\rm diag}(m_{i^0}),\quad
 U_L^{-- \dagger} \mathcal{M}_{--} U_R^{--} = {\rm diag}(m_{i^{--}}).
\end{align}

After decoupling the extra leptons, whose masses are typically given by the vectorlike mass $M_i$, they contribute to low-energy observables through higher dimensional operators in the SMEFT. 
They are represented as
\begin{align}
 \mathcal{L}_{d=6} &= \sum_j C_j \mathcal{O}_j,
\end{align}
where the dimension-six operators relevant for the current study are
\begin{align}
 \mathcal{O}_{e\phi} &= (\phi^\dagger\phi)(\bar\ell \phi \mu_R), \\
 \mathcal{O}_{\phi\ell}^{(1)} &= 
 (\phi^\dagger i \olrarrow{D_\mu} \phi) (\bar\ell \gamma^\mu \ell), \\
 \mathcal{O}_{\phi\ell}^{(3)} &= 
 (\phi^\dagger i \olrarrow{D_\mu^a} \phi) (\bar\ell \gamma^\mu \sigma^a \ell), \\
 \mathcal{O}_{\phi e} &= 
 (\phi^\dagger i \olrarrow{D_\mu} \phi) (\bar \mu_R \gamma^\mu \mu_R).
\end{align}
Note that $\ell$ denotes the lepton doublet in the second generation. 
Here, the derivatives mean
\begin{align}
 \phi^\dagger \olrarrow{D_\mu} \phi =
 \phi^\dagger (D_\mu \phi) - (D_\mu \phi)^\dagger \phi,~~~~~
 \phi^\dagger \olrarrow{D_\mu^a} \phi =
 \phi^\dagger \sigma^a (D_\mu \phi) - (D_\mu \phi)^\dagger \sigma^a \phi.
\end{align}
The Wilson coefficients are obtained as~\cite{delAguila:2008pw,deBlas:2017xtg}
\begin{align}
 C_{e\phi} &= y_\mu \left[
 \frac{\lambda_E^2}{2M_E^2} 
 + \frac{\lambda_{\Delta_1}^2}{2M_{\Delta_1}^2}
 + \frac{\lambda_{\Delta_3}^2}{2M_{\Delta_3}^2}
 + \frac{\lambda_{\Sigma_1}^2}{2M_{\Sigma_1}^2}
 \right] 
 \notag \\ &\quad
 - \frac{\lambda_E\lambda_{E\Delta_1}\lambda_{\Delta_1}}{M_E M_{\Delta_1}}
 - \frac{\lambda_E\lambda_{E\Delta_3}\lambda_{\Delta_3}}{M_E M_{\Delta_3}}
 - \frac{\lambda_{\Sigma_1}\lambda_{\Sigma_1\Delta_1}\lambda_{\Delta_1}}{M_{\Sigma_1} M_{\Delta_1}}
 + \frac{\lambda_{\Sigma_1}\lambda_{\Sigma_1\Delta_3}\lambda_{\Delta_3}}{M_{\Sigma_1} M_{\Delta_3}}, 
 \label{eq:CEH} \\
 C_{\phi\ell}^{(1)} &= 
 - \frac{\lambda_E^2}{4M_E^2}
 - \frac{3\lambda_{\Sigma_1}^2}{4M_{\Sigma_1}^2}, \\
 C_{\phi\ell}^{(3)} &= 
 - \frac{\lambda_E^2}{4M_E^2}
 + \frac{\lambda_{\Sigma_1}^2}{4M_{\Sigma_1}^2}, 
 \label{eq:CHE3} \\
 C_{\phi e} &= 
   \frac{\lambda_{\Delta_1}^2}{2M_{\Delta_1}^2}
 - \frac{\lambda_{\Delta_3}^2}{2M_{\Delta_3}^2},
\end{align}
at the tree level.
These coefficients are matched at the vectorlike mass scale.
In the following analysis, we ignore renormazliation group corrections below this scale for simplicity, which are induced by the SU(2)$_L$ and U(1)$_Y$ gauge interactions as well as the Yukawa couplings.

We define dimensionless coefficients as 
\begin{align}
 \widehat C_i = v^2\, C_i.
\end{align}
After the EWSB, the above operators modify the interactions of the Higgs, $W$ and $Z$ bosons from the SM predictions and affect low-energy observables. 
We will explain them in the following sections.

\section{Electroweak precision observables}
\label{sec:EWPO}

The Wilson coefficients $\widehat C_{e\phi}$, $\widehat C_{\phi\ell}^{(1)}$ and $\widehat C_{\phi\ell}^{(3)}$
are constrained strongly by the measurements of the EWPO, \ie,\ the $Z$ and $W$ boson observables (see Refs.~\cite{Efrati:2015eaa,Falkowski:2019hvp} for flavor-dependent studies). The EWPO can be calculated with the SM input parameters: the Fermi constant $G_F$, the fine structure constant $\alpha$, the strong coupling constant $\alpha_s(M_Z^2)$, 
the hadronic contribution $\Delta\alpha_{\mathrm{had}}^{(5)}(M_Z^2)$ 
to the renormalization-group running of $\alpha$, the $Z$-boson mass $M_Z$, 
the Higgs-boson masss $m_h$, the top-quark pole mass $m_t$, 
and other SM fermion masses. 
The measured values of the input parameters and the EWPO considered in this study are summarized in Table~\ref{tab:EWPO}.\footnote{The value of $\alpha_s(M_Z^2)$ is a lattice average calculated by the Flavour Lattice Averaging Group (FLAG)~\cite{Aoki:2019cca}. Also, the data of $m_t$ is found in the review section on {\it ``Electroweak Model and Constraints on New Physics"} of Ref.~\cite{Tanabashi:2018oca}.} In our numerical analysis, the parameters $G_F$, $\alpha$ and the light fermion masses are fixed to be constants~\cite{Tanabashi:2018oca}. 
\begin{table}[t]
\centering
\begin{tabular}{ccc|ccc}
\hline
& Measurement & Ref. &
& Measurement & Ref.
\\
\hline
$\alpha_s(M_Z^2)$ & 
$0.1182 \pm 0.0008$ &
\cite{Aoki:2019cca} &
$M_Z$ [GeV] &
$91.1876 \pm 0.0021$ &
\cite{Janot:2019oyi} 
\\
\cline{1-3}
$\Delta\alpha_{\mathrm{had}}^{(5)}(M_Z^2)$ &
$0.027609 \pm 0.000112$ &
\cite{Keshavarzi:2019abf} &
$\Gamma_Z$ [GeV] &
$2.4955 \pm 0.0023$ &
\\
\cline{1-3}
$m_t$ [GeV] &
$172.74 \pm 0.46$ &
\cite{Tanabashi:2018oca} &
$\sigma_{h}^{0}$ [nb] &
$41.4807 \pm 0.0325$ &
\\
\cline{1-3}
$m_h$ [GeV] &
$125.10 \pm 0.14$ &
\cite{Tanabashi:2018oca} &
$R^{0}_{e}$ &
$20.8038 \pm 0.0497$ &
\\
\cline{1-3}
$M_W$ [GeV] &
$80.379 \pm 0.012$ &
\cite{Tanabashi:2018oca} &
$R^{0}_{\mu}$ &
$20.7842 \pm 0.0335$ &
\\
\cline{1-3}
$\Gamma_{W}$ [GeV] & 
$2.085 \pm 0.042$ &
\cite{Tanabashi:2018oca} &
$R^{0}_{\tau}$ &
$20.7644 \pm 0.0448$ &
\\
\cline{1-3}
$\mathcal{B}(W\to e\nu)$ &
$0.1071 \pm 0.0016$ &
\cite{Schael:2013ita} &
$A_{\mathrm{FB}}^{0, e}$ &
$0.0145 \pm 0.0025$ &
\\
$\mathcal{B}(W\to \mu\nu)$ &
$0.1063 \pm 0.0015$ &
&
$A_{\mathrm{FB}}^{0, \mu}$ &
$0.0169 \pm 0.0013$ &
\\
$\mathcal{B}(W\to \tau\nu)$ &
$0.1138 \pm 0.002$ &
&
$A_{\mathrm{FB}}^{0, \tau}$ &
$0.0188 \pm 0.0017$ &
\\
\hline
$\mathcal{A}_e$ (SLD) & 
$ 0.1516 \pm 0.0021 $ &
\cite{ALEPH:2005ab} &
$R^{0}_{b}$ &  
$0.21629 \pm 0.00066$ & 
\cite{ALEPH:2005ab}
\\
$\mathcal{A}_\mu$ (SLD) & 
$ 0.142 \pm 0.015 $ &
&
$R^{0}_{c}$ & 
$0.1721 \pm 0.0030$ & 
\\
$\mathcal{A}_\tau$ (SLD) & 
$ 0.136 \pm 0.015 $ &
&
$A_{\mathrm{FB}}^{0, b}$ & 
$0.0992 \pm 0.0016$ &
\\
\cline{1-3}
$\mathcal{A}_e$ (LEP) & 
$ 0.1498 \pm 0.0049 $ &
\cite{ALEPH:2005ab}
&
$A_{\mathrm{FB}}^{0, c}$ & 
$0.0707 \pm 0.0035$ &
\\
$\mathcal{A}_\tau$ (LEP) & 
$ 0.1439 \pm 0.0043 $ &
&
$\mathcal{A}_b$ & 
$0.923 \pm 0.020$ &
\\
&
&
&
$\mathcal{A}_c$ & 
$0.670 \pm 0.027$ &
\\
\hline
\end{tabular}
\caption{Experimental measurement of the SM input parameters and EWPO.}
\label{tab:EWPO}
\end{table}

The operator $\mathcal{O}_{\phi\ell}^{(3)}$ alters the charged-current interactions of muon after the EWSB. Therefore the measured value of the Fermi constant $G_F$ from the muon decay involves a contribution from $\mathcal{O}_{\phi\ell}^{(3)}$: 
\begin{align}
 G_F 
 =
 \frac{1}{\sqrt{2}\,v^2}
 \left( 1 + \widehat{C}_{\phi\ell}^{(3)} \right)
 =
 \frac{1}{\sqrt{2}\,v^2}
 \left( 1 + \delta_{G_F} \right), 
 \label{eq:GF}
\end{align}
where $\delta_{G_F}\equiv \widehat{C}_{\phi\ell}^{(3)}$. The modification of $G_F$ affects the 
$W$-boson mass as 
\begin{align}
 m_W = (m_W)_{\rm SM} 
 \left[ 1 - \frac{s_W^2}{2(c_W^2-s_W^2)} \delta_{G_F} \right]. 
\end{align}
Here and hereafter, a quantity with the subscript ``SM'' denotes the corresponding SM prediction calculated with the measured values of the input parameters $G_F$, $\alpha$, $M_Z$, {\em etc}. The $W$-boson partial widths, which receive the corrections to $M_W$ and those to 
the charged-current couplings, are given by 
\begin{align}
 \Gamma(W^+ \to \mu^+\nu_\mu)
 &= 
 \Gamma(W^+ \to \mu^+\nu_\mu)_{\mathrm{SM}}
 \left[
 1 - \frac{1+c_W^2}{2(c_W^2-s_W^2)}\, \delta_{G_F} 
 + 2\,\widehat{C}_{\phi\ell}^{(3)}
 \right], \\
 \Gamma(W^+ \to ij)
 &=
 \Gamma(W^+ \to ij)_{\mathrm{SM}}
 \left[
 1 - \frac{1+c_W^2}{2(c_W^2-s_W^2)}\, \delta_{G_F}
 \right].
\end{align}
where $ij$ represents other final states including $e^+\nu_e$, $\tau^+\tau_\nu$ $\bar{d}u$ and $\bar{s}c$.

The operator $\mathcal{O}_{\phi\ell}^{(3)}$ also affects the neutral-current interactions of left-handed muon and muon neutrino. In addition, the operators $\mathcal{O}_{\phi\ell}^{(1)}$ and $\mathcal{O}_{\phi e}$ modify the neutral-current interactions. 
Taking account of the NP contribution in $G_F$, the $Z$-boson couplings to the SM fermions $f$ are modified as 
\begin{align}
 \mathcal{L}_Z
 &=
 \frac{g}{c_W}\, 
 \bar{f} \gamma^\mu 
 \Big[
 (T_L^{\prime 3} - Q s_W^2 + \delta g_L) P_L
 + 
 (T_R^{\prime 3} - Q s_W^2 + \delta g_R) P_R  \Big] f\,Z_\mu, 
\end{align}
where the corrections $\delta g_L$ and $\delta g_R$ are given by 
\begin{align}
 \delta g_L &=
 \begin{cases}
 \displaystyle
 - \frac{1}{2}
   \left[ T_L^{\prime 3} + \frac{Q s_W^2}{c_W^2-s_W^2} \right] \delta_{G_F}
 - \frac{1}{2}\,\widehat{C}_{\phi \ell}^{(1)}
 + T_L^{\prime 3}\, \widehat{C}_{\phi \ell}^{(3)}
 & \mathrm{for}~~f=\nu_L, \mu_L,
 \\[10pt]
\displaystyle
 - \frac{1}{2}
   \left[ T_L^{\prime 3} + \frac{Q s_W^2}{c_W^2-s_W^2} \right] \delta_{G_F}
 & \mathrm{otherwise},
 \end{cases}
 \\[5pt]
 \delta g_R &=
 \begin{cases}
 \displaystyle
 - \frac{Q s_W^2}{2(c_W^2-s_W^2)}\, \delta_{G_F}
 - \frac{1}{2}\,\widehat{C}_{\phi e}
 & \mathrm{for}~~f=\mu_R,
 \\[10pt]
 \displaystyle
 - \frac{Q s_W^2}{2(c_W^2-s_W^2)}\, \delta_{G_F}
 & \mathrm{otherwise}.
 \end{cases}
\end{align}
The $Z$-boson observables in Table~\ref{tab:EWPO} are written in terms of the effective $Zff$ couplings as shown, \eg,\ in Ref.~\cite{Ciuchini:2013pca}.

We perform a Bayesian fit of the Yukawa couplings $\lambda_{\Delta_1}$, $\lambda_{\Delta_3}$ and $\lambda_{\Sigma_1}$ to the experimental data of the EWPO, taking their correlations into account~\cite{ALEPH:2005ab,Schael:2013ita,Janot:2019oyi}. 
The $Z$-pole data at the LEP experiments have been updated recently in Ref.~\cite{Janot:2019oyi}, based on a sophisticated calculation of the Bhabha cross section, including beam-induced effects~\cite{Voutsinas:2019hwu}, for the measurement of the integrated luminosity. 
The fit is carried out with the \texttt{HEPfit} package~\cite{deBlas:2019okz}, which is based on the Markov Chain Monte Carlo provided by the Bayesian Analysis Toolkit (\texttt{BAT})~\cite{Caldwell:2008fw}. 
The SM contributions to $M_W$ and the $Z$-boson observables are calculated with the full  two-loop EW corrections using the approximate formulae presented in Refs.~\cite{Awramik:2003rn,Awramik:2006uz,Dubovyk:2019szj}, while the $W$-boson widths are calculated at one-loop level~\cite{Bardin:1986fi,Denner:1990tx}.

\section{Higgs decay}
\label{sec:Higgs}

The Higgs interactions are affected by the extra leptons through the SMEFT operators.
The muon Yukawa interaction is affected by $\mathcal{O}_{e\phi}$ as
\begin{align}
 \mathcal{L}_{\rm Yukawa} &= 
 -y_\mu \bar \ell \phi \mu_R + C_{e\phi}(\phi^\dagger\phi)(\bar \ell \phi \mu_R),
\end{align}
and thus, we obtain
\begin{align}
 y_\mu 
 = \sqrt{2}\,\frac{m_\mu}{v} + \frac{1}{2}\, \widehat{C}_{e\phi},
 = (y_\mu)_{\rm SM} \left[ 1 - \frac{1}{2}\, \delta_{G_F} \right]
 + \frac{1}{2}\, \widehat{C}_{e\phi},
 \label{eq:muYukawa}
\end{align}
after the EWSB.
Then, the Yukawa interaction is rewritten as
\begin{align}
 \mathcal{L}_{\rm Yukawa} &= 
 - m_\mu \bar \mu_L \mu_R  
 - \frac{1}{\sqrt{2}}\, (y_\mu)_{\rm SM}
 \left[ 
 1 - \frac{1}{2}\, \delta_{G_F}
 - \frac{1}{(y_\mu)_{\rm SM}}\, \widehat{C}_{e\phi}
 \right] h \bar \mu_L \mu_R + \cdots,
\end{align}
where $2h$ or $3h$ interactions are omitted. 
Consequently, the signal strength of the Higgs decay rate into muon pair is modified from the SM prediction as
\begin{align}
 \mu^{\mu\mu} 
 \equiv \frac{\Gamma(h \to \mu\mu)}{\Gamma(h \to \mu\mu)_{\rm SM}}
 = \left| 
 1 - \frac{1}{2}\, \delta_{G_F}
 - \frac{1}{(y_\mu)_{\rm SM}}\, \widehat{C}_{e\phi}
 \right|^2.
\end{align}
Experimentally, only the upper limits are set at 95\% CL as $\mu^{\mu\mu} < 2.1$ by ATLAS\cite{ATLAS:2018kbw} and $< 2.9$ by CMS~\cite{Sirunyan:2018hbu}.\footnote{
To be exact, the upper limits are imposed on $\sigma(pp\to h)\times B(h\to\mu\mu)/\sigma(pp\to h)_{\rm SM}\times B(h\to\mu\mu)_{\rm SM}$.
However, corrections of the extra leptons to the production cross section and the total decay rate of the Higgs boson are smaller by $\delta_{G_F}$ than the SM values, and thus, can be ignored safely.
}

\section{CKM unitarity}
\label{sec:CKM}

In the current setup, although the unitarity of the CKM matrix is maintained, the extra leptons can affect extractions of the CKM elements from experimental data.
In determining $|V_{ud}|$ from the superallowed $0^+\to 0^+$ nuclear $\beta$ decays, its transition rate is influenced by the extra leptons.
For example, the decay rate of a $\beta$ decay, $u \to d e^+ \nu$ is represented as
\begin{align}
 \Gamma_\beta 
 \propto \frac{1}{v^4}\, |V_{ud}|^2 
 = 2 G_F^2\, |V_{ud}|^2 \left( 1 + \widehat{C}_{\phi\ell}^{(3)} \right)^{-2},
\end{align}
via $\delta_{G_F}$, where we used Eq.~\eqref{eq:GF} in the last equality. 
Thus, by taking the EW radiative corrections into account, the superallowed $\beta$ decays satisfy the relation, (cf.~Ref.~\cite{Hardy:2018zsb})
\begin{align}
 |V_{ud}|^2 &= \left( 1 + \widehat{C}_{\phi\ell}^{(3)} \right)^2
 \frac{K}{2 \mathcal{F}t\, G_F^2 (1 + \Delta_R^V)}
 \notag \\ &=
 \left( 1 + \widehat{C}_{\phi\ell}^{(3)} \right)^2 \times
 \begin{cases}
 (0.97370 \pm 0.00014)^2 & \text{(SGPR)}, \\
 (0.97389 \pm 0.00018)^2 & \text{(CMS)}, \\
 (0.97365 \pm 0.00015)^2 & \text{(SFGJ)},
 \end{cases}
 \label{eq:Vud}
\end{align}
where $K = 8120.2776(9) \times 10^{-10}\GeV^{-4}s$ and $\mathcal{F}t = 3072.07(63)s$.\footnote{
 There may be additional nuclear corrections to $\mathcal{F}t$, which can introduce extra uncertainties~\cite{Seng:2018qru,Gorchtein:2018fxl}.
}
Also, the EW radiative corrections are given as
\begin{align}
\Delta_R^V
= 
\begin{cases}
0.02467 \pm 0.00022 & \text{(SGPR)~\cite{Seng:2018yzq}}, \\
0.02426 \pm 0.00032 & \text{(CMS)~\cite{Czarnecki:2019mwq}}, \\
0.02477 \pm 0.00024 & \text{(SFGJ)~\cite{Seng:2020wjq}}. 
\end{cases}
\end{align}

The CKM element $|V_{us}|$ is determined by measuring the $K$ meson decays.
A ratio $|V_{us}/V_{ud}|$ is extracted from 
a ratio of the leptonic decay rates of the $K$ and $\pi$ mesons~\cite{Passemar:2019}:
\begin{align}
 \left|\frac{V_{us}}{V_{ud}}\right|^2 = 
 \frac{\Gamma(K_{\mu2(\gamma)})}{\Gamma(\pi_{\mu2(\gamma)})}
 \frac{f_\pi^2}{f_K^2} 
 \frac{m_{\pi^\pm}(1-m_\mu^2/m_{\pi^\pm}^2)^2}{m_{K^\pm}(1-m_\mu^2/m_{K^\pm}^2)^2}
 ( 1-\delta ) 
 =
 (0.23129 \pm 0.00045)^2.
  \label{eq:VusKmu2}
\end{align}
Here, $f_K/f_\pi$ is a ratio of the $K$ and $\pi$ meson decay constants in the isospin limit, where the lattice results with $N_f=2+1+1$ are adopted.
The term $\delta$ takes account of EW radiative corrections and isospin-breaking effects.
It is noticed that Eq.~\eqref{eq:VusKmu2} is independent of the extra lepton contributions because the decay rates depend on $\widehat{C}_{\phi\ell}^{(3)}$ as~\cite{Coutinho:2019aiy}
\begin{align}
 \Gamma(K_{\mu2(\gamma)},\pi_{\mu2(\gamma)})
 \propto \frac{1}{v^4}\, |V_{us,ud}|^2 \left( 1 + \widehat{C}_{\phi\ell}^{(3)} \right)^2
 = 2 G_F^2\, |V_{us,ud}|^2, 
\end{align}
where Eq.~\eqref{eq:GF} is employed. 

The semileptonic $K$ meson decay rates are also used to determine $|V_{us}|$.
Their dependence on $\widehat{C}_{\phi\ell}^{(3)}$ are found as~\cite{Coutinho:2019aiy,Crivellin:2020lzu}
\begin{align}
 \Gamma(K_{e3})
 &\propto \frac{1}{v^4}\, |V_{us}|^2 
 = 2 G_F^2\, |V_{us}|^2 \left( 1 + \widehat{C}_{\phi\ell}^{(3)} \right)^{-2},
 \\
 \Gamma(K_{\mu3})
 &\propto \frac{1}{v^4}\, |V_{us}|^2 \left( 1 + \widehat{C}_{\phi\ell}^{(3)} \right)^2
 = 2 G_F^2\, |V_{us}|^2.
\end{align}
Hence, $|V_{us}|$ satisfies the relation, 
\begin{align}
 |V_{us}| = |V_{us}^{K_{e3}}| \left( 1 + \widehat{C}_{\phi\ell}^{(3)} \right),~~~
 |V_{us}| = |V_{us}^{K_{\mu3}}|,
\end{align}
where $|V_{us}^{K_{e3},K_{\mu3}}|$ are obtained by ignoring the extra lepton contributions, \ie, evaluated in the SM.
They are estimated as
\begin{align}
 |V_{us}^{K_{e3}}|   = 0.22320 \pm 0.00062,~~~
 |V_{us}^{K_{\mu3}}| = 0.22345 \pm 0.00068,
\end{align}
where the input values are summarized in Ref.~\cite{Moulson:2017ive,Cirigliano:2019}.
In particular, the form factor $f_+(0) = 0.9698(18)$ is obtained by lattice calculations with $N_f=2+1+1$~\cite{Carrasco:2016kpy,Bazavov:2018kjg}.\footnote{In Ref.~\cite{Cirigliano:2019}, the systematic uncertainty in the FNAL/MILC~18 result is taken to be 0.0011, but it has been updated to 0.0012 in the published version of the FNAL/MILC paper~\cite{Bazavov:2018kjg}. 
Accordingly, the uncertainty in $f_+(0)$ changes from 0.0017 to 0.0018.}

The above CKM elements satisfy the first-row CKM unitarity,
\begin{align}
 |V_{ud}|^2 + |V_{us}|^2 + |V_{ub}|^2 = 1,
\end{align}
which gives a constraint on $\widehat{C}_{\phi\ell}^{(3)}$ as
\begin{align}
 \widehat{C}_{\phi\ell}^{(3)} = 
 \begin{cases}
 (5.9 \pm 1.8) \times 10^{-4} 
 & (\text{SGPR},~K_{\mu2}/\pi_{\mu2}), \\
 (10.4 \pm 2.0) \times 10^{-4} 
 & (\text{SGPR},~K_{e3}), \\
 (10.4 \pm 2.2) \times 10^{-4} 
 & (\text{SGPR},~K_{\mu3}), \\
 (3.9 \pm 2.1) \times 10^{-4} 
 & (\text{CMS},~K_{\mu2}/\pi_{\mu2}), \\
 (8.5 \pm 2.3) \times 10^{-4} 
 & (\text{CMS},~K_{e3}), \\
 (8.4 \pm 2.5) \times 10^{-4} 
 & (\text{CMS},~K_{\mu3}), \\
 (6.4 \pm 1.8) \times 10^{-4} 
 & (\text{SFGJ},~K_{\mu2}/\pi_{\mu2}), \\
 (10.9 \pm 2.0) \times 10^{-4} 
 & (\text{SFGJ},~K_{e3}), \\
 (10.9 \pm 2.2) \times 10^{-4} 
 & (\text{SFGJ},~K_{\mu3}), \\
 \end{cases}
 \label{eq:CKM}
\end{align}
where $|V_{ub}| = 0.003683(75)$ is used~\cite{Charles:2004jd,CKMfitter:2019summer}.

In similar to the nuclear $\beta$ decays, the decays of the $\pi$ meson or the $\tau$ lepton are sensitive to the deviations of the $W$ boson interactions from the SM predictions. 
In the current setup, the extra leptons couple only to the muons, and thus, violate the LFU between $\pi \to \mu\nu$ and $\pi \to e\nu$ or between $\tau \to \mu\nu\bar\nu$ and $\tau \to e\nu\bar\nu$.
Although these decay modes give constraints on $\widehat{C}_{\phi\ell}^{(3)}$, the experimental uncertainties~\cite{Tanabashi:2018oca,Aguilar-Arevalo:2015cdf,Czapek:1993kc,Britton:1992pg,Amhis:2019ckw} are still large, and the constraints are weaker.

\section{Muon \texorpdfstring{$\boldsymbol{g-2}$}{g-2}}
\label{sec:gmin2}

The muon $g-2$ receives corrections from the extra leptons which couple to the muons.
In the mass eigenstate basis, the Higgs and gauge interactions are represented as
\begin{align}
 \mathcal{L}_{\rm int} &= 
 - \frac{1}{\sqrt{2}}\, g^{Hij}\, 
 \bar \psi^-_{Li} \psi^-_{Rj} h 
 + \frac{g}{c_W}\, g_{L,R}^{Zij}\, 
 \bar \psi^-_i \gamma^{\mu} P_{L,R} \psi^-_j Z_{\mu}
 \notag \\ &\quad
 + \frac{g}{\sqrt{2}}\, g_{L,R}^{W^1ij}\,
 \bar \psi^0_i \gamma^{\mu} P_{L,R} \psi^-_j W^+_{\mu}
 + \frac{g}{\sqrt{2}}\, g_{L,R}^{W^2ij}\, 
 \bar \psi^{--}_i \gamma^{\mu} P_{L,R} \psi^-_j W^-_{\mu}
 + \rm{h.c.},
\end{align}
where the couplings are given by
\begin{align}
 g^{Hij} &= \sum_{f,g} 
 (U_{L}^{-\,\dagger})_{if} (Y_-)_{fg} (U^-_{R})_{gj}, \label{eq:couplingH}\\
 g_{L,R}^{Zij} &= \sum_{f} 
 (U_{L,R}^{-\,\dagger})_{if} (T_{L,R}^{\prime 3} -s_W^2 Q)_f (U_{L,R}^-)_{fj}, \\
 g_{L,R}^{W^1ij} &= \sum_{f} 
 (U_{L,R}^{\prime 0\,\dagger})_{if}
 (U_{L,R}^-)_{fj} 
 \times
 \begin{cases}
 1 & \mbox{for}~f = \ell, \Delta_{1L}, \Delta_{1R}, \\
 \sqrt{2} & \mbox{for}~f = \Sigma_{1L}, \Sigma_{1R}, \\
 \end{cases} \\
 g_{L,R}^{W^2ij} &= \sum_{f} 
 (U_{L,R}^{\prime --\,\dagger})_{if}
 (U_{L,R}^-)_{fj} 
 \times
 \begin{cases}
 1 & \mbox{for}~f = \Delta_{3L}, \Delta_{3R}, \\
 \sqrt{2} & \mbox{for}~f = \Sigma_{1L}, \Sigma_{1R}. \\
 \end{cases}
 \label{eq:couplingW2}
\end{align}
Here, the fields indexed by $i,j$ are in the mass eigenstate basis, and $f, g$ in $g^{Hij}$ represent $\mu, E, \Delta_{1,3}^-$, and $\Sigma_1$ in the model basis. Also, $f$ in $g^{Zij}$ represent all the fields in Table \ref{tab:particle}.
Those in $g^{W^1ij}$ run over the fields which include the charge-neutral component, $\ell$, $\Delta_{1L,1R}$ and $\Sigma_1$. 
Similarly, $f$ in $g^{W^2ij}$ is effective for $\Delta_{3L,3R}$ and  $\Sigma_1$ and vanishing for the others. 

\begin{figure}[t]
\centering
\begin{subfigure}[b]{0.23\textwidth}
  \includegraphics[width=1.0\textwidth]{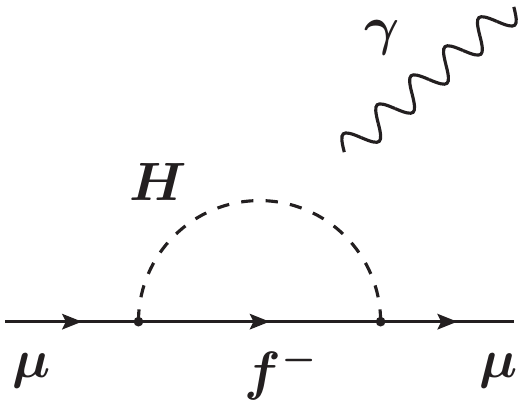}
 \end{subfigure}
 \hspace{0.01\textwidth}
 \begin{subfigure}[b]{0.23\textwidth}
  \includegraphics[width=1.0\textwidth]{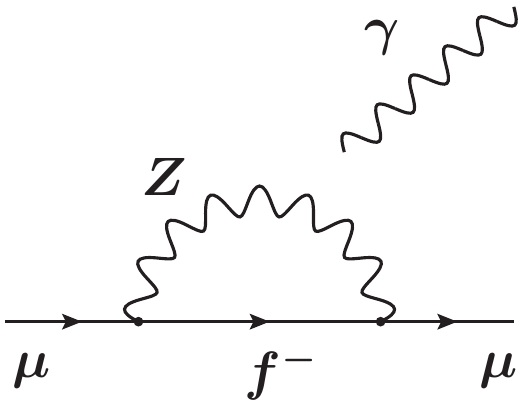}
 \end{subfigure}
 \hspace{0.01\textwidth}
 \begin{subfigure}[b]{0.23\textwidth}
  \includegraphics[width=1.0\textwidth]{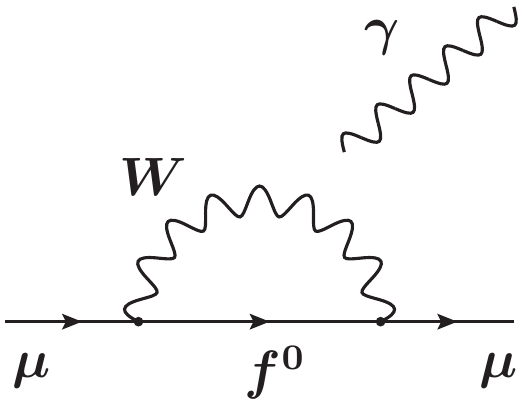}
 \end{subfigure}
 \hspace{0.01\textwidth}
 \begin{subfigure}[b]{0.23\textwidth}
  \includegraphics[width=1.0\textwidth]{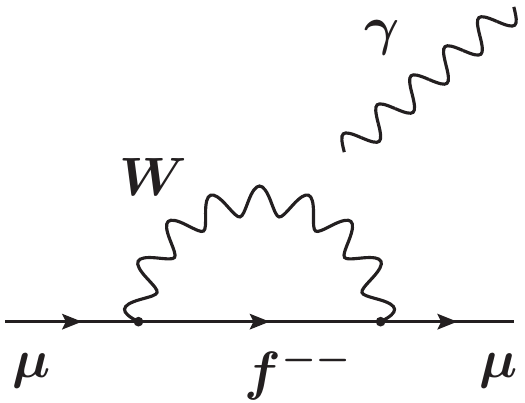}
 \end{subfigure}
 \caption{One-loop diagrams that contribute to the muon $g-2$, where $f^-$, $f^0$ and $f^{--}$ are extra leptons, and the photon attaches to charged particles.}
 \label{fig:g-2}
\end{figure}

As shown in Fig.~\ref{fig:g-2}, 
the loop diagrams for the muon $g-2$ are provided by exchanging the Higgs boson and singly-charged fermions for $a_{\mu}^{H}$, the $Z$ boson and singly-charged fermions for $a_{\mu}^{Z}$, the $W$ boson and neutrally-charged fermions for $a_{\mu}^{W^1}$, and the $W$ boson with doubly-charged fermions for $a_{\mu}^{W^2}$. 
The formulae for those contributions are found in Ref.~\cite{Freitas:2014pua} for the $W$ and $Z$ loop diagrams, while the reference \cite{Dermisek:2013gta} is used for the Higgs one. 
The results of the extra lepton contributions are summarized as
\begin{align}
 a_\mu^{\rm EL} = a_{\mu}^{H} + a_{\mu}^{Z} + a_{\mu}^{W^1} + a_{\mu}^{W^2},
\end{align}
where
\begin{align}
 a_{\mu}^{H} &= 
 \frac{m_{\mu}^2}{32\pi^2 m_H^2}
 \sum_{f^- \neq \mu}
 \biggl[ 
  \left[ (g^{Hf^-1})^2 + (g^{H1f^-})^2 \right]
  F_{\rm FFS} ( x_{f^-h} )
  + 
  g^{Hf^-1} g^{H1f^-}\, \frac{m_{f^-}}{m_{\mu}}\, 
  G_{\rm FFS} ( x_{f^-h} ) 
 \biggr],
 \label{eq:aH}\\
 a_{\mu}^{Z} &= 
 \frac{m_{\mu}^2 G_F}{2\sqrt{2}\pi^2}
 \sum_{f^- \neq \mu}
 \biggl[ 
  \left[ (g_L^{Zf^-1})^2 + (g_R^{Zf^-1})^2 \right]
  F_{\rm FFV} ( x_{f^-Z} )
  + g_L^{Zf^-1} g_R^{Zf^-1}\, \frac{m_{f^-}}{m_{\mu}}\, 
  G_{\rm FFV} ( x_{f^-Z} ) 
 \biggr],
 \\
 a_{\mu}^{W^1} &= 
 \frac{m_{\mu}^2 G_F}{4\sqrt{2}\pi^2}
 \sum_{f^0 \neq \nu}
 \biggl[ 
  \left[ (g_L^{W^1f^01})^2 + (g_R^{W^1f^01})^2 \right]
  F_{\rm VVF} ( x_{f^0W} ) 
 + g_L^{W^1f^01} g_R^{W^1f^01}\, \frac{m_{f^0}}{m_{\mu}}\, 
  G_{\rm VVF} ( x_{f^0W} ) 
 \biggr],
 \\
 a_{\mu}^{W^2} &= 
 \frac{m_{\mu}^2 G_F}{4\sqrt{2}\pi^2}
 \sum_{f^{--}}
 \biggl[ 
 \left[ (g_L^{W^2f^{--}1})^2 + (g_R^{W^2f^{--}1})^2 \right] 
 \left\{
  2 F_{\rm FFV} ( x_{f^{--}W} ) 
  - F_{\rm VVF} ( x_{f^{--}W} ) 
 \right\}
 \notag \\ & \qquad\qquad\qquad\qquad
 + 
 g_L^{W^2f^{--}1} g_R^{W^2f^{--}1}\,\frac{m_{f^{--}}}{m_{\mu}}
 \left\{
  2 G_{\rm FFV} ( x_{f^{--}W} ) 
  - G_{\rm VVF} ( x_{f^{--}W} ) 
 \right\}
 \biggr]
 \label{eq:aW2}
\end{align}
with $x_{ij} = m_i^2/m_j^2$.
Here, the unitary matrices $U_i$ in Eqs.~\eqref{eq:couplingH}--\eqref{eq:couplingW2} are defined such that the mass eigenstates are ordered from lightest to heaviest, and thus, ``1'' in the indices of the coupling constants in Eqs.~\eqref{eq:aH}--\eqref{eq:aW2} means the muon-like fermion in the mass eigenstate basis. 
The loop functions are defined as~\cite{Freitas:2014pua}
\begin{align}
 F_{\rm FFS}(x) &= \frac{1}{6(x-1)^4} \bigl[x^3-6x^2+3x+2+6x \ln x\bigr], \\
 G_{\rm FFS}(x) &= \frac{1}{(x-1)^3} \bigl[x^2 - 4x + 3 + 2 \ln x\bigr], \\
 F_{\rm FFV}(x) &= \frac{1}{6(x-1)^4} \bigl[-5x^4+14x^3-39x^2+38x-8+18x^2 \ln x\bigr], \\
 G_{\rm FFV}(x) &= \frac{1}{(x-1)^3} \bigl[x^3+3x-4-6x\ln x\bigr], \\
 F_{\rm VVF}(x) &= \frac{1}{6(x-1)^4} \bigl[4x^4-49x^3+78x^2-43x+10+18x^3 \ln x\bigr],\\
 G_{\rm VVF}(x) &= \frac{1}{(x-1)^3} \bigl[-x^3+12x^2-15x+4-6x^2 \ln x\bigr].
\end{align}

All of the extra lepton contributions, Eqs.~\eqref{eq:aH}--\eqref{eq:aW2}, can be enhanced by $\lambda_i/y_\mu$ where $\lambda_i = \lambda_{E\Delta_1}$, $\lambda_{E\Delta_3}$, $\lambda_{\Sigma_1\Delta_1}$, and $\lambda_{\Sigma_1\Delta_3}$.
In fact, any contribution to the muon $g-2$ involves a chirality flippling on the fermion line, and it is provided by these Yukawa couplings rather than the muon one. 
Consequently, $\Delta a_{\mu}$ is approximated as
\begin{align}
 a_\mu^{\rm EL} = \sum_{i,j}
 \delta_{ij} \frac{v^2 \lambda_{i}\lambda_{ij} \lambda_{j}}{M_{i}M_{j}},
 \label{eq:gmin2approx}
\end{align}
for $M_E \sim M_{\Delta_1} \sim M_{\Delta_3} \sim M_{\sigma_1} \gg v$.
The coefficients are estimated as $\delta_{ij} \sim -2 \times 10^{-6}$, $-1 \times 10^{-5}$, $-2 \times 10^{-6}$, and $2 \times 10^{-6}$ for $(i,j) = (E,\Delta_1)$, $(E,\Delta_3)$, $(\Sigma_1,\Delta_1)$, and $(\Sigma_1,\Delta_3)$.
It is noticed that the sign of each contribution is determined by $\lambda_{ij}$.
Besides, the Yukawa couplings $\lambda_{\Delta_1E}$, $\lambda_{\Delta_3E}$, $\lambda_{\Delta_1\Sigma_1}$, $\lambda_{\Delta_3\Sigma_1}$ do not affect the muon $g-2$ significantly.

The contributions that are not chirally enhanced are safely negligible in the limit of $M_i \gg v$.
In particular, we do not include extra contributions from the SM loop diagrams, \ie, $f^-, f^0 \neq 1$ in Eqs.~\eqref{eq:aH}--\eqref{eq:aW2}.
The SM Higgs, $Z$ and $W$ coupling constants are modified by the extra leptons via the unitary matrices $U_i$. 
Such deviations induce extra contributions by exchanging the SM particles in the loop diagrams. 
However, they are not chirally enhanced, and thus, ignored in the analysis.

\section{Result}
\label{sec:result}

\begin{figure}[t]
 \centering
 \includegraphics[width=0.5\textwidth]{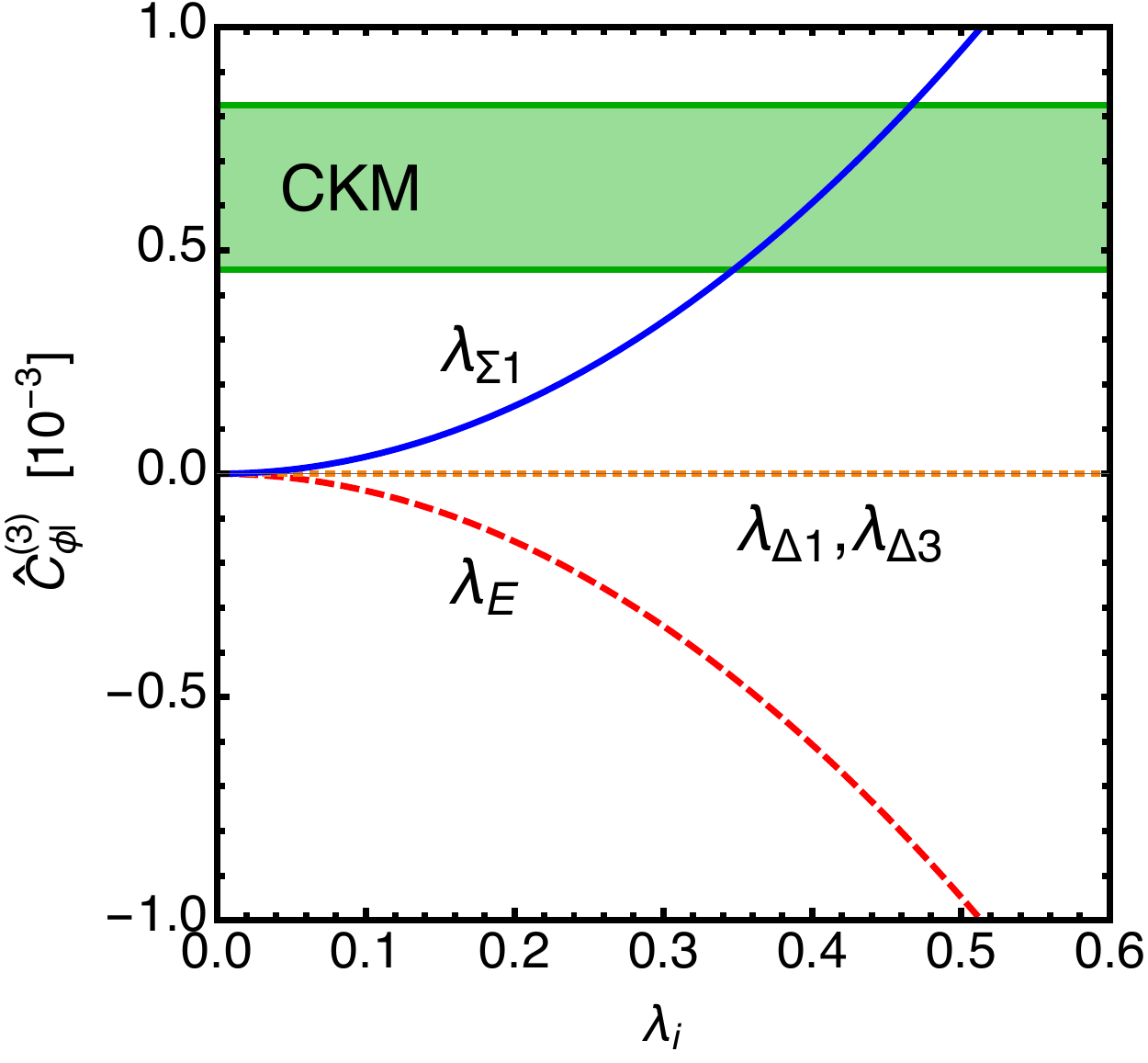}
 \caption{$\widehat C_{\phi\ell}^{(3)}$ in the extra lepton models with the vectorlike masses $M_i = 2\TeV$. In the green band, $\widehat C_{\phi\ell}^{(3)}$ favored by the CKM unitarity is explained at the $1\sigma$ level, where the SFGJ value and the decay rates of $K_{\mu2},\pi_{\mu2}$ are used.}
\label{fig:CHE}
\end{figure}

First of all, let us study the CKM unitarity in the extra lepton models. 
As we explained in Sec.~\ref{sec:CKM}, $|V_{ud}|$ determined by the nuclear $\beta$ decays and $|V_{us}|$ by $K_{e3}$ are affected by $\widehat{C}_{\phi\ell}^{(3)}$, \ie, by $E$ and $\Sigma_1$ among the extra leptons.
The result is shown in Fig.~\ref{fig:CHE}, where $\widehat{C}_{\phi\ell}^{(3)}$ is plotted as functions of the Yukawa couplings $\lambda_i$. 
Here, the vectorlike masses are set to be $M_i = 2\TeV$.
The green band shows the $1\sigma$ region of $\widehat{C}_{\phi\ell}^{(3)}$ favored by the CKM unitarity, where the SFGJ result is adopted for $|V_{ud}|$ and the decay rates of $K_{\mu2},\pi_{\mu2}$ are used for $|V_{us}|$. 
It is found that only $\Sigma_1$ can relax the tension in the CKM unitarity.
Depending on the evaluations of $\Delta_R^V$ and $|V_{us}|$, the CKM unitarity favors the regions,
\begin{align}
 \lambda_{\Sigma_1} = 
 \begin{cases}
 0.40 \pm 0.06 
 & (\text{SGPR},~K_{\mu2}/\pi_{\mu2}), \\
 0.52 \pm 0.05 
 & (\text{SGPR},~K_{e3}), \\
 0.52_{-0.06}^{+0.05} 
 & (\text{SGPR},~K_{\mu3}), \\
 0.32_{-0.10}^{+0.08} 
 & (\text{CMS},~K_{\mu2}/\pi_{\mu2}), \\
 0.47_{-0.07}^{+0.06} 
 & (\text{CMS},~K_{e3}), \\
 0.47_{-0.07}^{+0.06} 
 & (\text{CMS},~K_{\mu3}), \\
 0.41 \pm 0.06
 & (\text{SFGJ},~K_{\mu2}/\pi_{\mu2}), \\
 0.54 \pm 0.05 
 & (\text{SFGJ},~K_{e3}), \\
 0.54_{-0.06}^{+0.05} 
 & (\text{SFGJ},~K_{\mu3}), \\
 \end{cases}
 \label{eq:lambdaSigma1}
\end{align}
at the $1\sigma$ level.
This result is scaled by a ratio $\lambda_{\Sigma_1}/M_{\Sigma_1}$ for $M_{\Sigma_1} \neq 2\TeV$ because $\widehat{C}_{\phi\ell}^{(3)}$ is proportional to the ratio squared.
On the other hand, since the extra lepton $E$ decreases $\widehat C_{\phi\ell}^{(3)}$, its contribution is favored to be decoupled by suppressing $\lambda_E$ or assuming $M_E \gg v$. 
In the following analysis, we assume $\lambda_E=0$.

\begin{figure}[p]
 \centering
 \begin{subfigure}[b]{0.48\textwidth}
  \includegraphics[width=\textwidth]{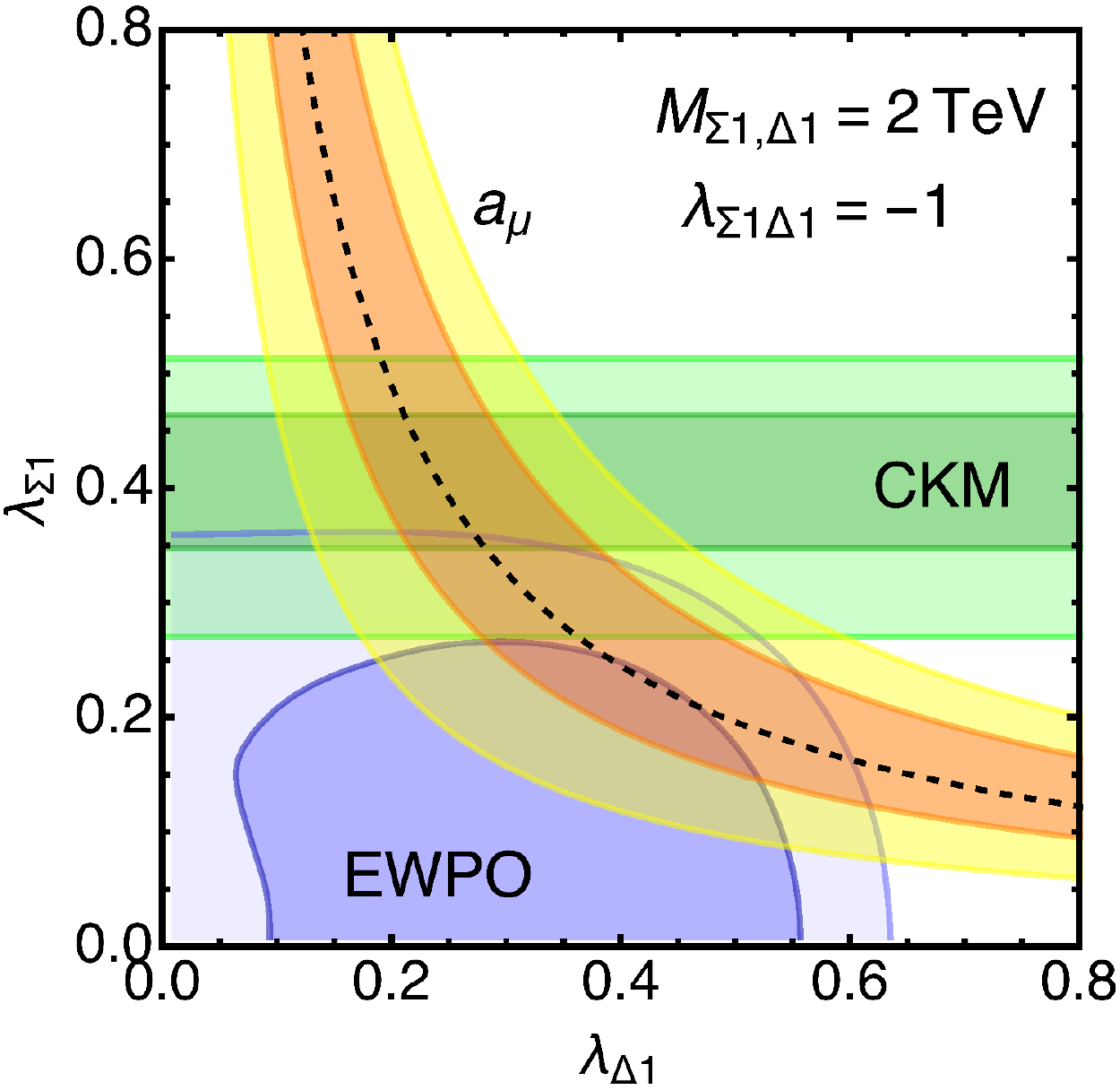}
  \vspace{.2cm}
 \end{subfigure}
 \begin{subfigure}[b]{0.48\textwidth}
  \includegraphics[width=\textwidth]{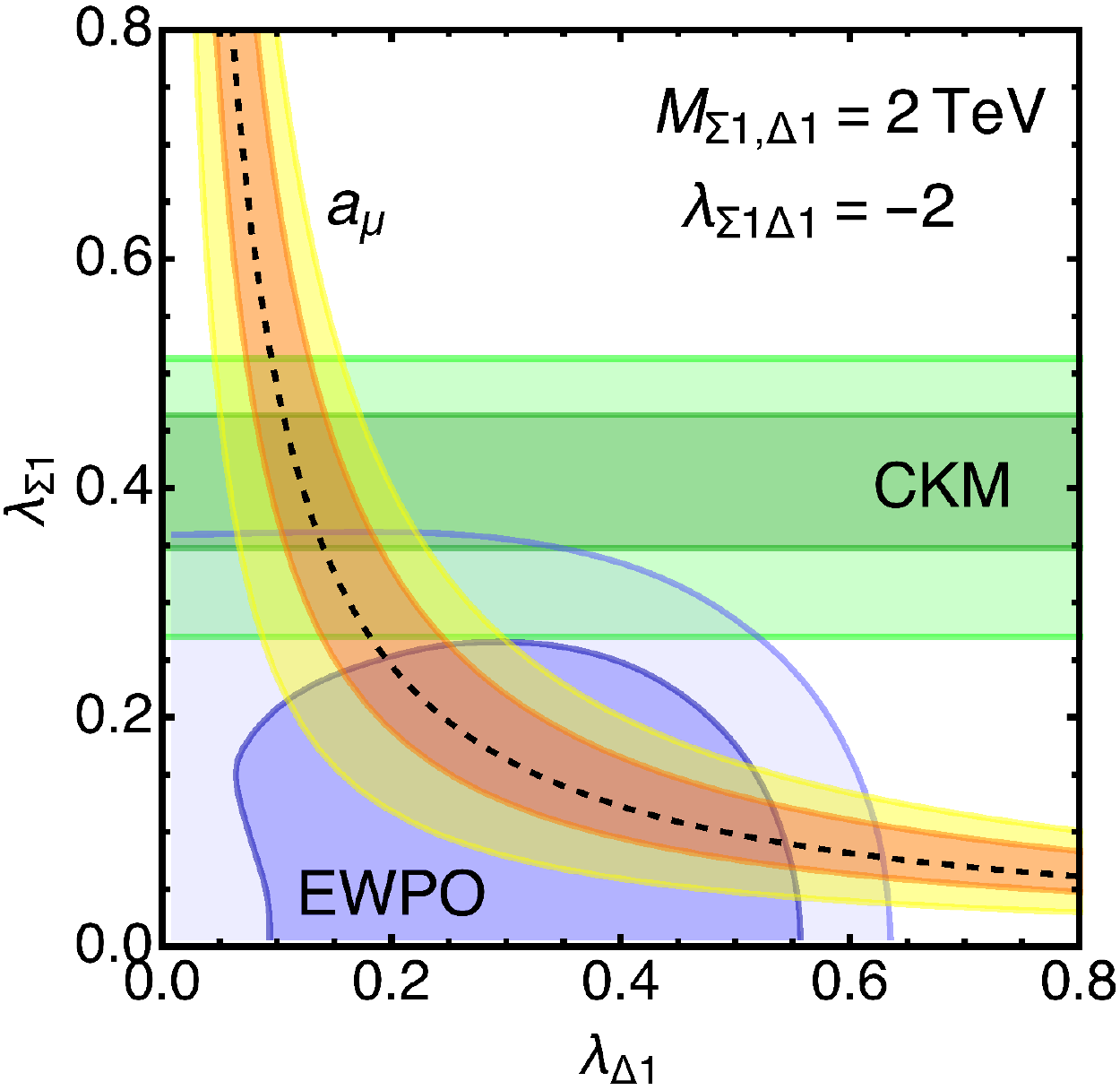}
  \vspace{.2cm}
 \end{subfigure}
 \begin{subfigure}[b]{0.48\textwidth}
  \includegraphics[width=\textwidth]{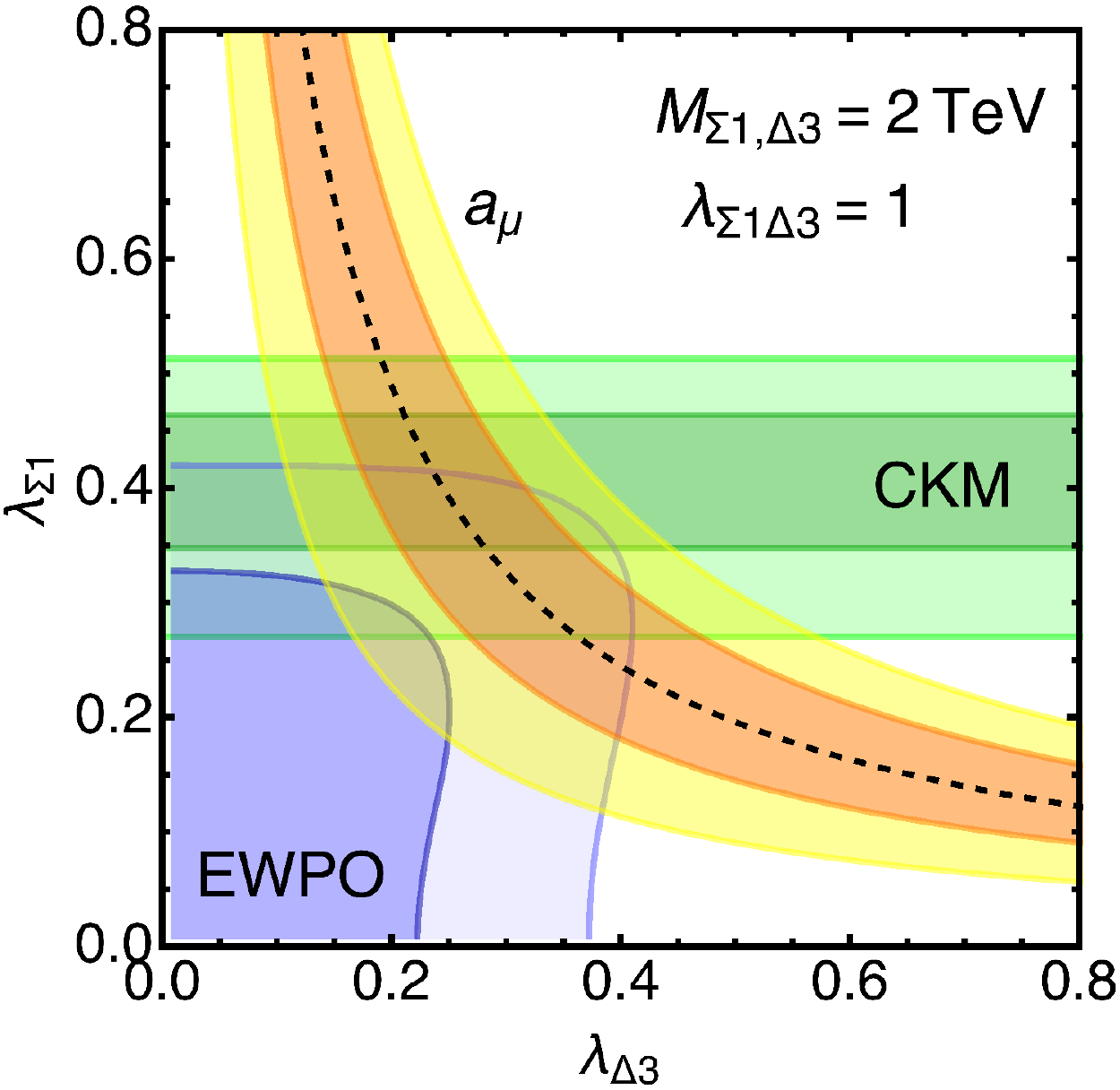}
 \end{subfigure}
 \begin{subfigure}[b]{0.48\textwidth}
  \includegraphics[width=\textwidth]{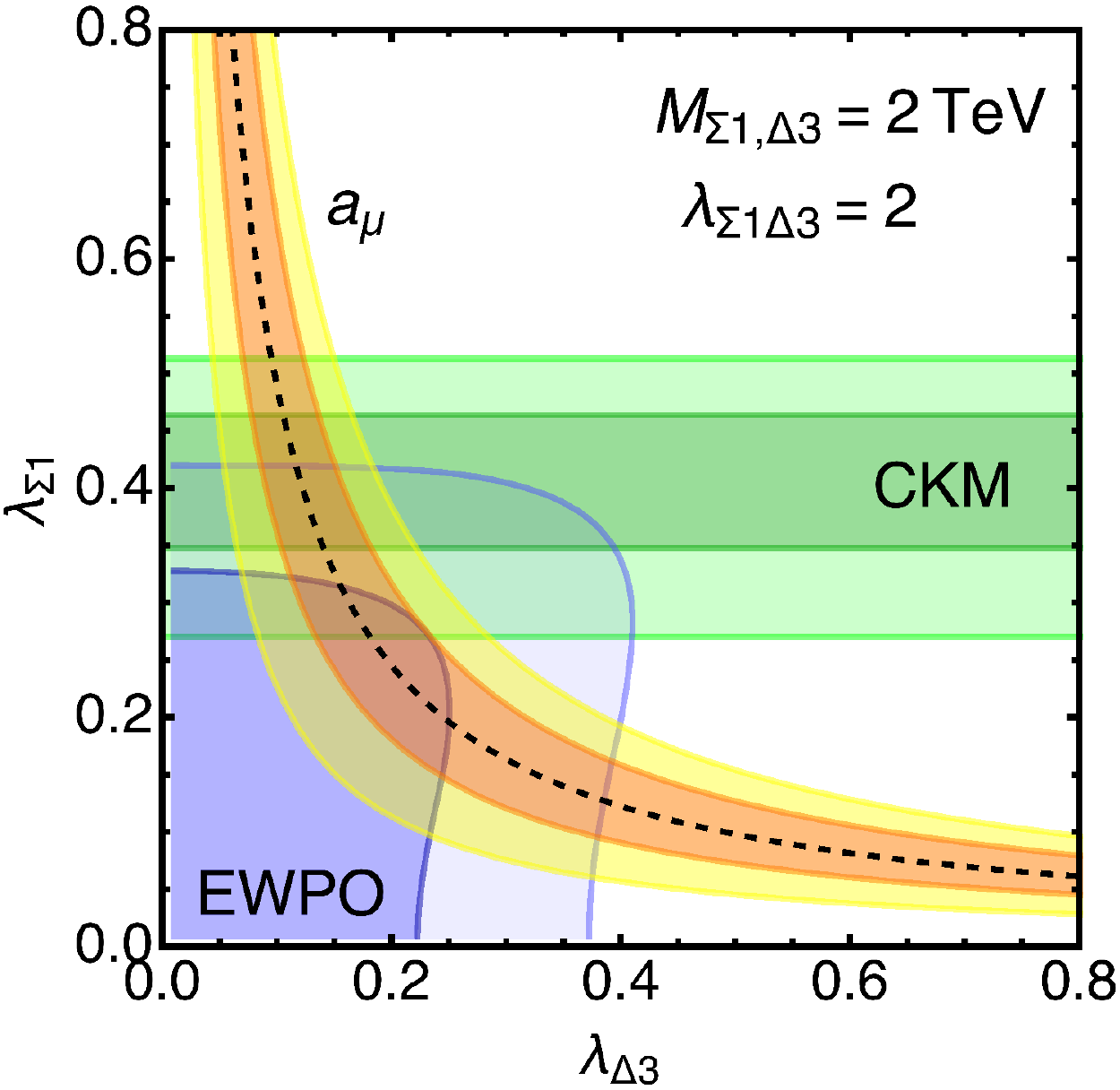}
 \end{subfigure}
\vspace{0.5cm}
\caption{
  The CKM elements (green), $a_\mu$ (orange, yellow), EWPO (blue) and $\mu^{\mu\mu}$ are evaluated as functions of $\lambda_{\Sigma_{1}}$ and $\lambda_{\Delta_{1,3}}$. 
  The other Yukawa couplings which are not shown explicitly in each plot are set to be zero. 
  For each observable, the current result is explained at the $1\sigma$ $(2\sigma)$ level in the thick (thin) colored region, while the upper limit from the Higgs signal strenght $\mu^{\mu\mu}$ is drawn by the black dashed line, where the left side is allowed at 95\% CL by ATLAS.
  Here, the SFGJ value and the decay rates of $K_{\mu2},\pi_{\mu2}$ are used to obtain the CKM regions. 
  }
\label{fig:summary}
\end{figure}

Next, let us consider the tension in the muon $g-2$.
According to Eq.~\eqref{eq:gmin2approx}, the contributions of $\Sigma_1$ can be chirally enhanced if it is accompanied by $\Delta_1$ or $\Delta_3$. 
Once $\lambda_{\Delta_1}$ or $\lambda_{\Delta_3}$ is turned on, EWPO is also affected via $C_{\phi e}$ in similar to $\lambda_{\Sigma_1}$ through $C_{\phi \ell}^{(1,3)}$. 
In Fig.~\ref{fig:summary}, the muon $g-2$ and EWPO as well as the CKM elements are evaluated as functions of the Yukawa couplings;
in the top (bottom) plots, $\lambda_{\Sigma_1}$ and $\lambda_{\Delta_1}$ $(\lambda_{\Delta_3})$ are turned on, while  $\lambda_{\Delta_3} = 0$ $(\lambda_{\Delta_1} = 0)$ is assumed. 
Here, all the vectorlike masses are set to be $M_i = 2\TeV$.
In the left (right) plots, $|\lambda_{\Sigma_1\Delta_{1,3}}| = 1$ $(2)$ is chosen, and its sign is determined such that the extra lepton contribution to the muon $g-2$ becomes positive. 
Also, since all the observables are insensitive to $\lambda_{\Delta_{1,3}\Sigma_1}$, it is set to be zero here and hereaftrer.
For each observable, the current data is explained at the $1\sigma$ $(2\sigma)$ level in the thick (thin) colored region.
Here, the SFGJ value is adopted for $|V_{ud}|$ and the decay rates of $K_{\mu2},\pi_{\mu2}$ are used for $|V_{us}|$. 

It is found that both of the tensions in the CKM unitarity and the muon $g-2$ can be solved under the constraint from EWPO for $|\lambda_{\Sigma_1\Delta_{1,3}}| = \mathcal{O}(1)$ at $M_i = 2\TeV$.
Since the extra lepton contributions to the CKM elements and EWPO are proportional to powers of $\lambda_i/M_i$, the corresponding parameter regions are simply scaled from Fig.~\ref{fig:summary} as $M_i$ is varied.
On the other hand, since those to the muon $g-2$ are scaled by $\lambda_{\Sigma_1}\lambda_{\Delta_{1,3}}/M_{\Sigma_1}M_{\Delta_{1,3}} \times \lambda_{\Sigma_1\Delta_{1,3}}$, the parameter region favored by the muon $g-2$ depends on $\lambda_{\Sigma_1\Delta_{1,3}}$ in the figure.

\begin{figure}[t]
 \centering
 \begin{subfigure}[b]{0.44\textwidth}
  \includegraphics[width=\textwidth]{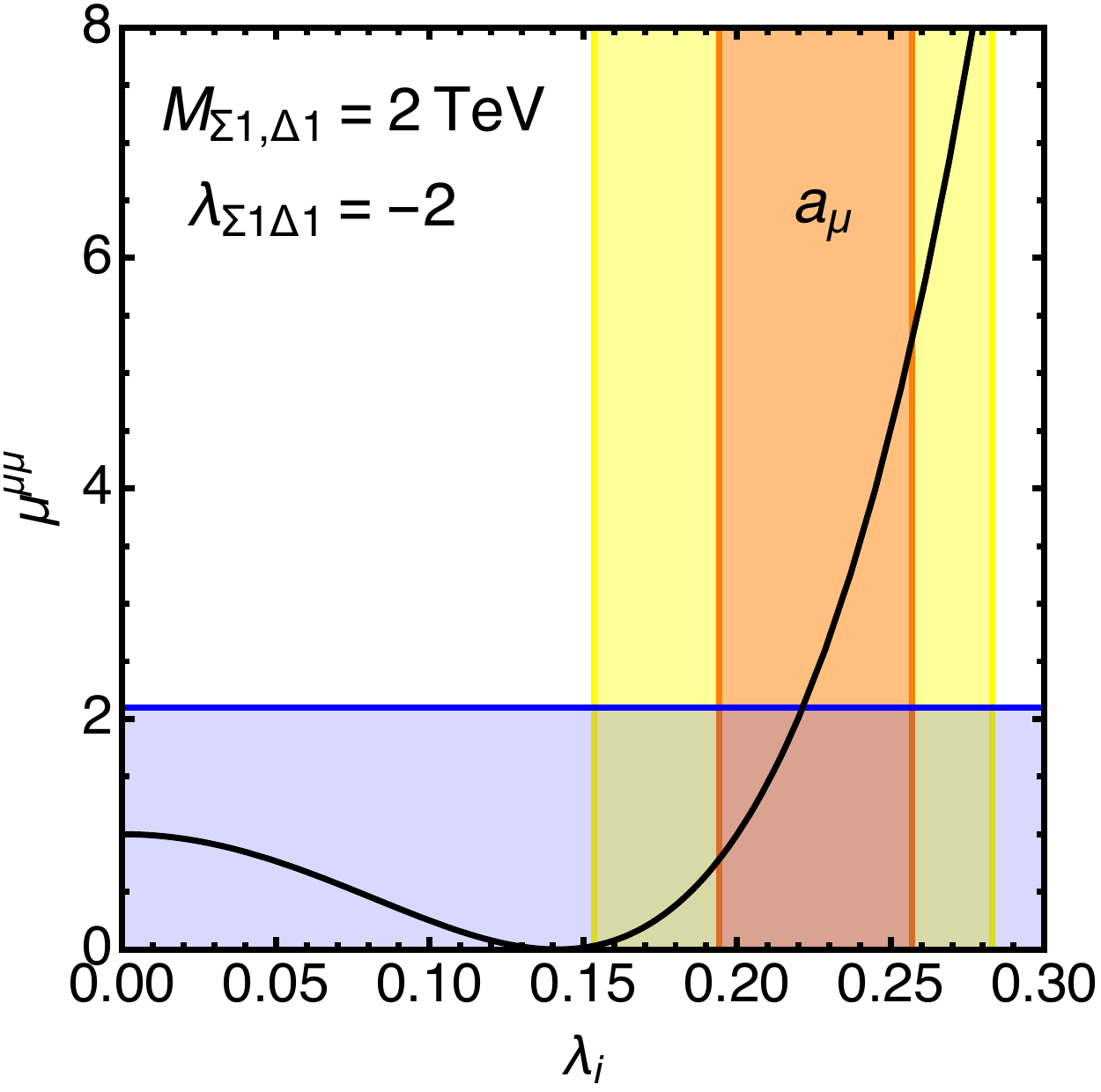}
 \end{subfigure}\hspace{5mm}
 \begin{subfigure}[b]{0.44\textwidth}
  \includegraphics[width=\textwidth]{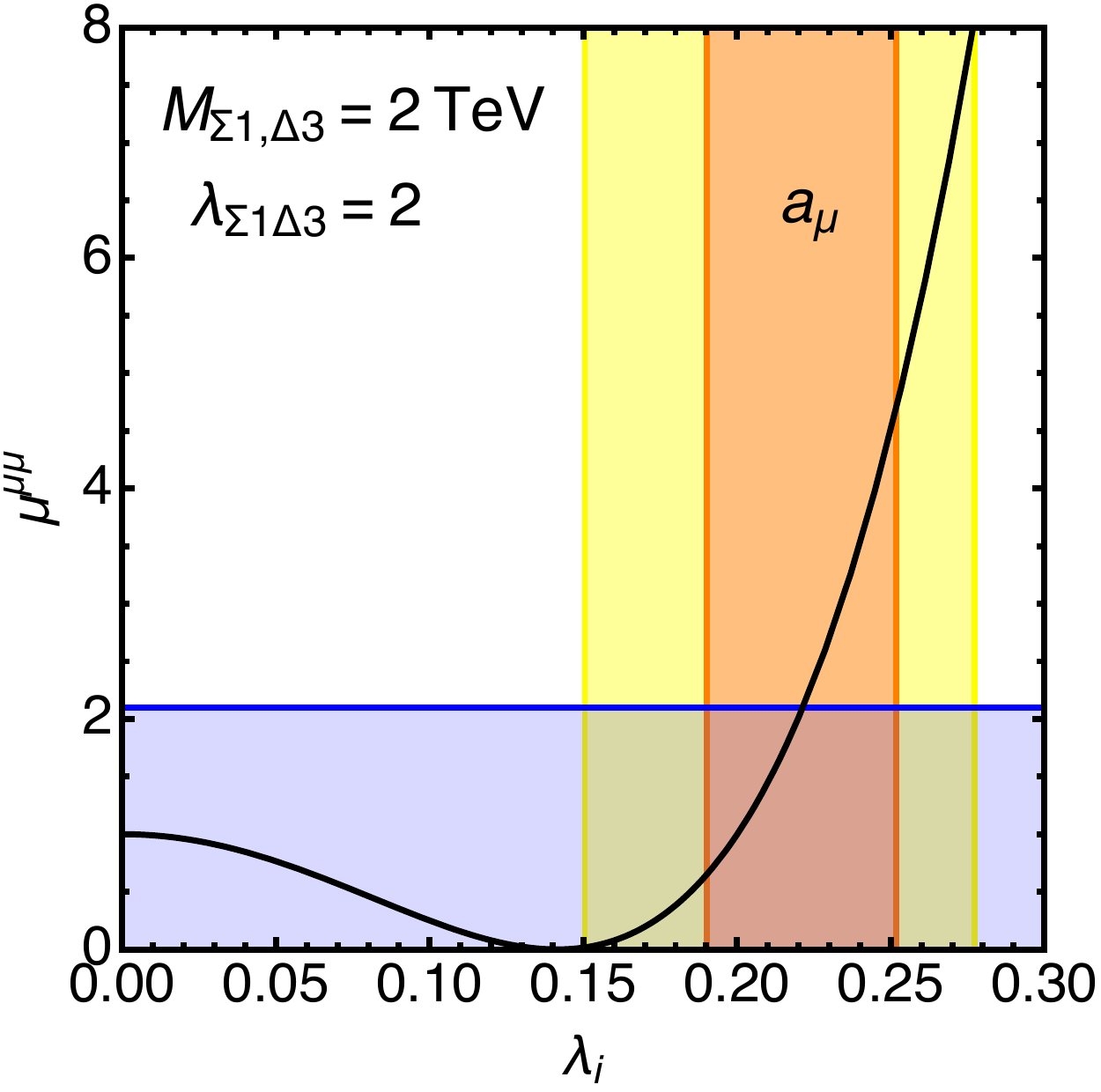}
 \end{subfigure}
\caption{
  Signal strength of $h \to \mu\mu$ as a function of the Yukawa coupling $\lambda_i$.
  Here, $\lambda_i \equiv \lambda_{\Sigma_1} = \lambda_{\Delta_1}$ with $\lambda_{\Delta_3} = 0$ (left) and $\lambda_i \equiv \lambda_{\Sigma_1} = \lambda_{\Delta_3}$ with $\lambda_{\Delta_1} = 0$ (right). 
  Also, $|\lambda_{\Sigma_1\Delta_{1,3}}| = 2$ and $M_i = 2\TeV$.
  The blue region is allowed at 95\% level by a search for $h \to \mu\mu$ at ATLAS.
  The discrepancy in the muon $g-2$ is explained at the $1\sigma$ $(2\sigma)$ level by the Yukawa couplings in the orange (yellow) region. 
  }
\label{fig:Hmumu}
\end{figure}

The correction to the muon Yukawa interaction \eqref{eq:muYukawa} is magnified if the extra lepton contribution to the muon $g-2$ is enhanced.
It is required to be as large as the SM value in the parameter region where the muon $g-2$ is explained.
Hence, tight parameter tunings between $y_\mu$ and the extra lepton contribution are not necessary to achieve the muon mass. 
However, such a contribution is limited by the Higgs decay rate into muon pair.
In Fig.~\ref{fig:Hmumu}, the signal strength of the Higgs decay rate $\mu^{\mu\mu}$ is shown as a function of the Yukawa coupling $\lambda_i \equiv \lambda_{\Sigma_1} = \lambda_{\Delta_{1,3}}$.
In the left plot, $\lambda_{\Delta_1}$ is turned on with $\lambda_{\Delta_3} = 0$, and  $\Delta_1 \leftrightarrow \Delta_3$ in the right plot. 
Here, $|\lambda_{\Sigma_1\Delta_{1,3}}| = 2$ and $M_i = 2\TeV$.
The blue region is allowed at 95\% level by ATLAS.
On the other hand, the discrepancy in the muon $g-2$ is explained at the $1\sigma$ $(2\sigma)$ level by the Yukawa couplings in the orange (yellow) region. 
It is found that almost a half of the muon $g-2$ parameter region is already excluded by $h \to \mu\mu$.

The constraint from $\mu^{\mu\mu}$ is also shown in Fig.~\ref{fig:summary}.
The left region of the black dashed line is allowed at 95\% level.
We conclude that both of the tensions in the CKM unitarity and the muon $g-2$ can be solved simultaneously under the constraints from $\mu^{\mu\mu}$ as well as the EWPO.
Note that since the correction to the Yukawa interaction \eqref{eq:CEH} is dominated by $\widehat{C}_{e\phi}$, its parameter dependence on $\lambda_i$ and $M_i$ is the same as that of the chirally-enhanced contribution to the muon $g-2$ \eqref{eq:gmin2approx}.
Thus, the above conclusion is insensitive to the choice of $\lambda_{\Sigma_1\Delta_{1,3}}$ and $M_i$.

As seen from Fig.~\ref{fig:Hmumu}, the experimental uncertainty of $\mu^{\mu\mu}$ as well as that of the muon $g-2$ is large. 
In particular, $\mu^{\mu\mu} = 1$, \ie, the SM value, is consistent with the current discrepancy of the muon $g-2$.
Thus, we cannot confirm/refute the extra lepton contribution yet, and it is significant to reduce both uncertainties.
In future, the HL-LHC experiment may achieve $\delta \mu^{\mu\mu}/\mu^{\mu\mu} = 9$\% at $\sqrt{s}=14\TeV$ with the integrated luminosity $\mathcal{L}=6$\,ab$^{-1}$, and the uncertainty could be reduced by an order of magnitude compared to HL-LHC at FCC-ee/eh/hh~\cite{deBlas:2019rxi}.
Also, the experimental uncertainty of the muon $g-2$ is planned to be reduced by a factor of 4 compared to the current value in the near future~\cite{Grange:2015fou,Keshavarzi:2019bjn,Mibe:2011zz,Abe:2019thb}.
Therefore, we expect to check the extra lepton model by these measurements in future.

The extra leptons also contribute to the decay rate of the Higgs boson into two photons. 
The corrections are induced by $\delta_{G_F}$ and the extra lepton loops.
They are estimated to be $\mathcal{O}(0.1)\%$ of the SM prediction, which is well within the current experimental uncertainty~\cite{Sirunyan:2018koj,Aad:2019mbh}.
In future, the experimental precision may become $\delta \mu^{\gamma\gamma}/\mu^{\gamma\gamma} = 3$\% at HL-LHC with $\sqrt{s}=14\TeV$ and $\mathcal{L}=6$\,ab$^{-1}$, and would be $0.6$\% at FCC-ee/eh/hh~\cite{deBlas:2019rxi}.
Thus, the extra lepton contribution to the two-photon channel could be probed in future. 

In Fig.~\ref{fig:summary}, the CKM elements are evaluated by adopting the SFGJ result and the decay rates of $K_{\mu2}$ and $\pi_{\mu2}$.
If we use the CMS evaluation for $\Delta_R^V$, the parameter overlapping with the EWPO region becomes better (see Eq.~\eqref{eq:lambdaSigma1} for a favored value of $\lambda_{\Sigma_1}$).
On the other hand, larger $\lambda_{\Sigma_1}$ is favored by $V_{us}$ determined by $K_{\ell3}$.
The CKM region becomes consistent with the EWPO constraint at the $2\sigma$ level if the CMS evaluation is adopted, while it is not the case for SFGJ or SGPR.
In any case, the tension between $V_{ud}$ determined by the nuclear $\beta$ decays and $V_{us}$ by the $K$ meson decay is relaxed by the extra lepton $\Sigma_1$.\footnote{
 According to Eq.~\eqref{eq:CKM}, it is noticed that the discrepancy between $V_{us}$ determined by $K_{\mu2},\pi_{\mu2}$ and that by $K_{\ell3}$ cannot be solved in the current framework.
}

Before closing this section, let us comment on the direct searches for the extra leptons. 
At collider experiments, they can be produced by exchanging the SM gauge bosons and decay predominantly into the SM bosons $W, Z, h$ and the muonic leptons $\mu, \nu$.
Such particles have signatures with multilepton final states. Although there are no experimental analyses based on the full dataset of LHC Run-II, the model may be excluded if the vectorlike masses are $M_i \sim 100\GeV$ (cf.~the CMS analysis~\cite{Sirunyan:2019ofn} for the tauonic extra lepton search at $\sqrt{s} = 13\TeV$, and Refs.~\cite{Aad:2015dha,Falkowski:2013jya,Dermisek:2014qca,Ellis:2014dza,Kumar:2015tna} based on the LHC result at $\sqrt{s} = 8\TeV$). 
Thus, the setup with $M_i = 2\TeV$ safely avoids the direct searches for the extra leptons at the LHC experiments.
On the other hand, since future proton-proton colliders such as HL-LHC or higher energy colliders have potentials to probe those particles in multi-TeV scales~\cite{Bhattiprolu:2019vdu}, the extra leptons which solve the tensions in the CKM unitarity and the muon $g-2$ could be discovered.\footnote{
Triplet lepton searches are also studied for future electron-positron and electron-proton colliders~\cite{Das:2020gnt}.
}

\section{Conclusions}
\label{sec:conclusion}

Motivated by the tensions reported in the CKM unitarity and the muon $g-2$, we studied the models of extra leptons which couple to the muon and have vectorlike masses. 
It was shown that the former tension is solved by introducing an SU(2)$_L$ triplet $\Sigma_1$.
In addition, the contribution to the muon $g-2$ can be enhanced if it is accompanied by an SU(2)$_L$ doublet $\Delta_1$ or $\Delta_3$. 
At the same time, the models are constrained by the EWPO and the Higgs boson decays. 
We found that both of the tensions can be solved simultaneously under these constraints. 
In particular, the Higgs decay rate into two muons is likely to be modified from the SM prediction significantly, and thus, could be useful to test the model at future experiments (see \eg, Ref.~\cite{deBlas:2019rxi}). 

The above tensions are planned to be checked in future. 
Prospects for the test of the CKM unitarity and the LFU violations are discussed in Ref.~\cite{Crivellin:2020lzu}. 
Also, the experimental value of the muon $g-2$ will be updated in the near future~\cite{Grange:2015fou,Keshavarzi:2019bjn,Mibe:2011zz,Abe:2019thb}.
Once the tensions would be confirmed, the extra lepton models can provide one of the attractive scenarios.

\section*{Acknowledgements}
This work is supported in part by the Grant-in-Aid for
Scientific Research B (No.16H03991 [ME]), 
Early-Career Scientists (No.16K17681 [ME]) and 
Scientific Research C (No.17K05429 [SM]). 

\bibliographystyle{utphys28mod}
\bibliography{references}

\providecommand{\href}[2]{#2}\begingroup\raggedright\begin{thebibliography}{10}

\bibitem{Aoyama:2020ynm}
T.~Aoyama {\em et~al.}, {\em {The anomalous magnetic moment of the muon in the
  Standard Model}.} {\ttfamily
  \href{https://arxiv.org/abs/2006.04822}{arXiv:2006.04822}}.

\bibitem{Borsanyi:2020mff}
S.~Borsanyi {\em et~al.}, {\em {Leading-order hadronic vacuum polarization
  contribution to the muon magnetic momentfrom lattice QCD}.} {\ttfamily
  \href{https://arxiv.org/abs/2002.12347}{arXiv:2002.12347}}.

\bibitem{Davier:2019can}
M.~Davier, A.~Hoecker, B.~Malaescu, and Z.~Zhang, {\em {A new evaluation of the
  hadronic vacuum polarisation contributions to the muon anomalous magnetic
  moment and to $\alpha(m_Z^2)$},}
  \href{https://dx.doi.org/10.1140/epjc/s10052-020-7792-2}{Eur.\  Phys.\  J.\
  {\bfseries C80} (2020) 241}
{\ttfamily [\href{https://arxiv.org/abs/1908.00921}{arXiv:1908.00921}]}.

\bibitem{Keshavarzi:2019abf}
A.~Keshavarzi, D.~Nomura, and T.~Teubner, {\em {$g-2$ of charged leptons,
  $\alpha (M^2_Z)$ , and the hyperfine splitting of muonium},}
  \href{https://dx.doi.org/10.1103/PhysRevD.101.014029}{Phys.\  Rev.\
  {\bfseries D101} (2020) 014029}
{\ttfamily [\href{https://arxiv.org/abs/1911.00367}{arXiv:1911.00367}]}.

\bibitem{Bennett:2002jb}
{\bfseries Muon g-2} Collaboration, {\em {Measurement of the positive muon
  anomalous magnetic moment to 0.7 ppm},}
  \href{https://dx.doi.org/10.1103/PhysRevLett.89.129903,
  10.1103/PhysRevLett.89.101804}{Phys.\  Rev.\  Lett.\  {\bfseries 89} (2002)
  101804} {\ttfamily
  [\href{https://arxiv.org/abs/hep-ex/0208001}{hep-ex/0208001}]}.
[Erratum: Phys. Rev. Lett.89,129903(2002)].

\bibitem{Bennett:2004pv}
{\bfseries Muon g-2} Collaboration, {\em {Measurement of the negative muon
  anomalous magnetic moment to 0.7 ppm},}
  \href{https://dx.doi.org/10.1103/PhysRevLett.92.161802}{Phys.\  Rev.\  Lett.\
   {\bfseries 92} (2004) 161802}
{\ttfamily [\href{https://arxiv.org/abs/hep-ex/0401008}{hep-ex/0401008}]}.

\bibitem{Bennett:2006fi}
{\bfseries Muon g-2} Collaboration, {\em {Final Report of the Muon E821
  Anomalous Magnetic Moment Measurement at BNL},}
  \href{https://dx.doi.org/10.1103/PhysRevD.73.072003}{Phys.\  Rev.\
  {\bfseries D73} (2006) 072003}
{\ttfamily [\href{https://arxiv.org/abs/hep-ex/0602035}{hep-ex/0602035}]}.

\bibitem{CODATA2018}
E.~Tiesinga, P.~J.~Mohr, D.~B.~Newell, and B.~N.~Taylor, {\em {The 2018 CODATA
  Recommended Values of the Fundamental Physical Constants},} {Web Version 8.1,
  2020. Database developed by J.~Baker, M.~Douma, and S.~Kotochigova,
  \url{http://physics.nist.gov/constants}}.

\bibitem{Parker:2018vye}
R.~H.~Parker, C.~Yu, W.~Zhong, B.~Estey, and H.~Müller, {\em {Measurement of
  the fine-structure constant as a test of the Standard Model},}
  \href{https://dx.doi.org/10.1126/science.aap7706}{Science {\bfseries 360}
  (2018) 191} {\ttfamily
  [\href{https://arxiv.org/abs/1812.04130}{arXiv:1812.04130}]}.

\bibitem{Hardy:2014qxa}
J.~C.~Hardy and I.~S.~Towner, {\em {Superallowed $0^+\to 0^+$ nuclear $\beta$
  decays: 2014 critical survey, with precise results for $V_{ud}$ and CKM
  unitarity},} \href{https://dx.doi.org/10.1103/PhysRevC.91.025501}{Phys.\
  Rev.\  C {\bfseries 91} (2015) 025501} {\ttfamily
  [\href{https://arxiv.org/abs/1411.5987}{arXiv:1411.5987}]}.

\bibitem{Hardy:2016vhg}
J.~Hardy and I.~S.~Towner, {\em {$|V_{ud}|$ from nuclear $\beta$ decays},}
  \href{https://dx.doi.org/10.22323/1.291.0028}{PoS {\bfseries CKM2016} (2016)
  028}.

\bibitem{Hardy:2018zsb}
J.~C.~Hardy and I.~S.~Towner in {\em {13th Conference on the Intersections of
  Particle and Nuclear Physics (CIPANP 2018) Palm Springs, California, USA, May
  29-June 3, 2018}}.
\newblock 2018.
\newblock
{\ttfamily \href{https://arxiv.org/abs/1807.01146}{arXiv:1807.01146}}.
\newblock

\bibitem{Marciano:2005ec}
W.~J.~Marciano and A.~Sirlin, {\em {Improved calculation of electroweak
  radiative corrections and the value of V(ud)},}
  \href{https://dx.doi.org/10.1103/PhysRevLett.96.032002}{Phys.\  Rev.\  Lett.\
   {\bfseries 96} (2006) 032002} {\ttfamily
  [\href{https://arxiv.org/abs/hep-ph/0510099}{hep-ph/0510099}]}.

\bibitem{Seng:2018yzq}
C.-Y.~Seng, M.~Gorchtein, H.~H.~Patel, and M.~J.~Ramsey-Musolf, {\em {Reduced
  Hadronic Uncertainty in the Determination of $V_{ud}$},}
  \href{https://dx.doi.org/10.1103/PhysRevLett.121.241804}{Phys.\  Rev.\
  Lett.\  {\bfseries 121} (2018) 241804}
{\ttfamily [\href{https://arxiv.org/abs/1807.10197}{arXiv:1807.10197}]}.

\bibitem{Czarnecki:2019mwq}
A.~Czarnecki, W.~J.~Marciano, and A.~Sirlin, {\em {Radiative Corrections to
  Neutron and Nuclear Beta Decays Revisited},}
  \href{https://dx.doi.org/10.1103/PhysRevD.100.073008}{Phys.\  Rev.\
  {\bfseries D100} (2019) 073008}
{\ttfamily [\href{https://arxiv.org/abs/1907.06737}{arXiv:1907.06737}]}.

\bibitem{Seng:2020wjq}
C.-Y.~Seng, X.~Feng, M.~Gorchtein, and L.-C.~Jin, {\em {Joint lattice QCD -
  dispersion theory analysis confirms the top-row CKM unitarity deficit}.}
  {\ttfamily \href{https://arxiv.org/abs/2003.11264}{arXiv:2003.11264}}.

\bibitem{Cirigliano:2019}
V.~Cirigliano, M.~Moulson, and E.~Passemar, {\em The status of $V_{us}$,} Talk
  given at the workshop ``Current and Future Status of First-Row CKM
  Unitarity," UMass Amherst, May 2019,
  \url{https://www.physics.umass.edu/acfi/sites/acfi/files/slides/moulson_amherst.pdf}.

\bibitem{Passemar:2019}
E.~Passemar and M.~Moulson, {\em Extraction of $V_{us}$ from experimental
  measurements,} Talk given at the International Conference on Kaon Physics
  2019, Univ. of Perugia, Sep. 2019,
  \url{https://indico.cern.ch/event/769729/contributions/3512047/attachments/1905114/3146148/Kaon2019_MoulsonPassemarCorr.pdf}.

\bibitem{Belfatto:2019swo}
B.~Belfatto, R.~Beradze, and Z.~Berezhiani, {\em {The CKM unitarity problem: A
  trace of new physics at the TeV scale?}}
  \href{https://dx.doi.org/10.1140/epjc/s10052-020-7691-6}{Eur.\  Phys.\  J.\
  {\bfseries C80} (2020) 149}
{\ttfamily [\href{https://arxiv.org/abs/1906.02714}{arXiv:1906.02714}]}.

\bibitem{Grossman:2019bzp}
Y.~Grossman, E.~Passemar, and S.~Schacht, {\em {On the Statistical Treatment of
  the Cabibbo Angle Anomaly}.}
{\ttfamily \href{https://arxiv.org/abs/1911.07821}{arXiv:1911.07821}}.

\bibitem{Charles:2004jd}
{\bfseries CKMfitter Group} Collaboration, {\em {CP violation and the CKM
  matrix: Assessing the impact of the asymmetric $B$ factories},}
  \href{https://dx.doi.org/10.1140/epjc/s2005-02169-1}{Eur.\ \ Phys.\ \ J.\ \ C
  {\bfseries 41} (2005) 1--131} {\ttfamily
  [\href{https://arxiv.org/abs/hep-ph/0406184}{hep-ph/0406184}]}.

\bibitem{CKMfitter:2019summer}
{\bfseries CKMfitter Group} Collaboration,
  \url{http://ckmfitter.in2p3.fr/www/results/plots_summer19/num/ckmEval_results_summer19.html}.

\bibitem{Coutinho:2019aiy}
A.~M.~Coutinho, A.~Crivellin, and C.~A.~Manzari, {\em {Global Fit to Modified
  Neutrino Couplings and the Cabibbo-Angle Anomaly}.}
{\ttfamily \href{https://arxiv.org/abs/1912.08823}{arXiv:1912.08823}}.

\bibitem{Crivellin:2020lzu}
A.~Crivellin and M.~Hoferichter, {\em {Beta decays as sensitive probes of
  lepton flavor universality}.}
{\ttfamily \href{https://arxiv.org/abs/2002.07184}{arXiv:2002.07184}}.

\bibitem{delAguila:2008pw}
F.~del Aguila, J.~de~Blas, and M.~Perez-Victoria, {\em {Effects of new leptons
  in Electroweak Precision Data},}
  \href{https://dx.doi.org/10.1103/PhysRevD.78.013010}{Phys.\  Rev.\
  {\bfseries D78} (2008) 013010}
{\ttfamily [\href{https://arxiv.org/abs/0803.4008}{arXiv:0803.4008}]}.

\bibitem{Dermisek:2013gta}
R.~Dermisek and A.~Raval, {\em {Explanation of the Muon g-2 Anomaly with
  Vectorlike Leptons and its Implications for Higgs Decays},}
  \href{https://dx.doi.org/10.1103/PhysRevD.88.013017}{Phys.\  Rev.\
  {\bfseries D88} (2013) 013017}
{\ttfamily [\href{https://arxiv.org/abs/1305.3522}{arXiv:1305.3522}]}.

\bibitem{Czarnecki:2001pv}
A.~Czarnecki and W.~J.~Marciano, {\em {The Muon anomalous magnetic moment: A
  Harbinger for `new physics'},}
  \href{https://dx.doi.org/10.1103/PhysRevD.64.013014}{Phys.\ \ Rev.\ \ D
  {\bfseries 64} (2001) 013014} {\ttfamily
  [\href{https://arxiv.org/abs/hep-ph/0102122}{hep-ph/0102122}]}.

\bibitem{Kannike:2011ng}
K.~Kannike, M.~Raidal, D.~M.~Straub, and A.~Strumia, {\em {Anthropic solution
  to the magnetic muon anomaly: the charged see-saw},}
  \href{https://dx.doi.org/10.1007/JHEP02(2012)106}{JHEP {\bfseries 02} (2012)
  106} {\ttfamily [\href{https://arxiv.org/abs/1111.2551}{arXiv:1111.2551}]}.
  [Erratum: JHEP 10, 136 (2012)].

\bibitem{Freitas:2014pua}
A.~Freitas, J.~Lykken, S.~Kell, and S.~Westhoff, {\em {Testing the Muon g-2
  Anomaly at the LHC},} \href{https://dx.doi.org/10.1007/JHEP09(2014)155,
  10.1007/JHEP05(2014)145}{JHEP {\bfseries 05} (2014) 145} {\ttfamily
  [\href{https://arxiv.org/abs/1402.7065}{arXiv:1402.7065}]}.
[Erratum: JHEP09,155(2014)].

\bibitem{Megias:2017dzd}
E.~Megias, M.~Quiros, and L.~Salas, {\em {$g_\mu-2$ from Vector-Like Leptons in
  Warped Space},} \href{https://dx.doi.org/10.1007/JHEP05(2017)016}{JHEP
  {\bfseries 05} (2017) 016} {\ttfamily
  [\href{https://arxiv.org/abs/1701.05072}{arXiv:1701.05072}]}.

\bibitem{Poh:2017tfo}
Z.~Poh and S.~Raby, {\em {Vectorlike leptons: Muon g-2 anomaly, lepton flavor
  violation, Higgs boson decays, and lepton nonuniversality},}
  \href{https://dx.doi.org/10.1103/PhysRevD.96.015032}{Phys.\ \ Rev.\ \ D
  {\bfseries 96} (2017) 015032} {\ttfamily
  [\href{https://arxiv.org/abs/1705.07007}{arXiv:1705.07007}]}.

\bibitem{Kowalska:2017iqv}
K.~Kowalska and E.~M.~Sessolo, {\em {Expectations for the muon g-2 in
  simplified models with dark matter},}
  \href{https://dx.doi.org/10.1007/JHEP09(2017)112}{JHEP {\bfseries 09} (2017)
  112} {\ttfamily [\href{https://arxiv.org/abs/1707.00753}{arXiv:1707.00753}]}.

\bibitem{Raby:2017igl}
S.~Raby and A.~Trautner, {\em {Vectorlike chiral fourth family to explain muon
  anomalies},} \href{https://dx.doi.org/10.1103/PhysRevD.97.095006}{Phys.\ \
  Rev.\ \ D {\bfseries 97} (2018) 095006} {\ttfamily
  [\href{https://arxiv.org/abs/1712.09360}{arXiv:1712.09360}]}.

\bibitem{Calibbi:2018rzv}
L.~Calibbi, R.~Ziegler, and J.~Zupan, {\em {Minimal models for dark matter and
  the muon g$-$2 anomaly},}
  \href{https://dx.doi.org/10.1007/JHEP07(2018)046}{JHEP {\bfseries 07} (2018)
  046} {\ttfamily [\href{https://arxiv.org/abs/1804.00009}{arXiv:1804.00009}]}.

\bibitem{Crivellin:2018qmi}
A.~Crivellin, M.~Hoferichter, and P.~Schmidt-Wellenburg, {\em {Combined
  explanations of $(g-2)_{\mu,e}$ and implications for a large muon EDM},}
  \href{https://dx.doi.org/10.1103/PhysRevD.98.113002}{Phys.\ \ Rev.\ \ D
  {\bfseries 98} (2018) 113002} {\ttfamily
  [\href{https://arxiv.org/abs/1807.11484}{arXiv:1807.11484}]}.

\bibitem{Kawamura:2019rth}
J.~Kawamura, S.~Raby, and A.~Trautner, {\em {Complete vectorlike fourth family
  and new U(1)' for muon anomalies},}
  \href{https://dx.doi.org/10.1103/PhysRevD.100.055030}{Phys.\ \ Rev.\ \ D
  {\bfseries 100} (2019) 055030} {\ttfamily
  [\href{https://arxiv.org/abs/1906.11297}{arXiv:1906.11297}]}.

\bibitem{Kawamura:2019hxp}
J.~Kawamura, S.~Raby, and A.~Trautner, {\em {Complete vectorlike fourth family
  with U(1)' : A global analysis},}
  \href{https://dx.doi.org/10.1103/PhysRevD.101.035026}{Phys.\ \ Rev.\ \ D
  {\bfseries 101} (2020) 035026} {\ttfamily
  [\href{https://arxiv.org/abs/1911.11075}{arXiv:1911.11075}]}.

\bibitem{Minkowski:1977sc}
P.~Minkowski, {\em {$\mu \to e\gamma$ at a Rate of One Out of $10^{9}$ Muon
  Decays?}} \href{https://dx.doi.org/10.1016/0370-2693(77)90435-X}{Phys.\ \
  Lett.\ \ B {\bfseries 67} (1977) 421--428}.

\bibitem{GellMann:1980vs}
M.~Gell-Mann, P.~Ramond, and R.~Slansky, {\em {Complex Spinors and Unified
  Theories},} Conf.\ \ Proc.\ \ C {\bfseries 790927} (1979) 315--321 {\ttfamily
  [\href{https://arxiv.org/abs/1306.4669}{arXiv:1306.4669}]}.

\bibitem{Yanagida:1979as}
T.~Yanagida, {\em {Horizontal gauge symmetry and masses of neutrinos},} Conf.\
  \ Proc.\ \ C {\bfseries 7902131} (1979) 95--99.

\bibitem{Mohapatra:1979ia}
R.~N.~Mohapatra and G.~Senjanovic, {\em {Neutrino Mass and Spontaneous Parity
  Nonconservation},}
  \href{https://dx.doi.org/10.1103/PhysRevLett.44.912}{Phys.\ \ Rev.\ \ Lett.\
  {\bfseries 44} (1980) 912}.

\bibitem{Foot:1988aq}
R.~Foot, H.~Lew, X.~G.~He, and G.~C.~Joshi, {\em {Seesaw Neutrino Masses
  Induced by a Triplet of Leptons},}
\href{https://dx.doi.org/10.1007/BF01415558}{Z.\  Phys.\  {\bfseries C44}
  (1989) 441}.

\bibitem{deBlas:2017xtg}
J.~de~Blas, J.~C.~Criado, M.~Perez-Victoria, and J.~Santiago, {\em {Effective
  description of general extensions of the Standard Model: the complete
  tree-level dictionary},}
  \href{https://dx.doi.org/10.1007/JHEP03(2018)109}{JHEP {\bfseries 03} (2018)
  109}
{\ttfamily [\href{https://arxiv.org/abs/1711.10391}{arXiv:1711.10391}]}.

\bibitem{Efrati:2015eaa}
A.~Efrati, A.~Falkowski, and Y.~Soreq, {\em {Electroweak constraints on
  flavorful effective theories},}
  \href{https://dx.doi.org/10.1007/JHEP07(2015)018}{JHEP {\bfseries 07} (2015)
  018} {\ttfamily [\href{https://arxiv.org/abs/1503.07872}{arXiv:1503.07872}]}.

\bibitem{Falkowski:2019hvp}
A.~Falkowski and D.~Straub, {\em {Flavourful SMEFT likelihood for Higgs and
  electroweak data},} \href{https://dx.doi.org/10.1007/JHEP04(2020)066}{JHEP
  {\bfseries 04} (2020) 066} {\ttfamily
  [\href{https://arxiv.org/abs/1911.07866}{arXiv:1911.07866}]}.

\bibitem{Aoki:2019cca}
{\bfseries Flavour Lattice Averaging Group} Collaboration, {\em {FLAG Review
  2019},} \href{https://dx.doi.org/10.1140/epjc/s10052-019-7354-7}{Eur.\
  Phys.\  J.\  {\bfseries C80} (2020) 113}
{\ttfamily [\href{https://arxiv.org/abs/1902.08191}{arXiv:1902.08191}]}.

\bibitem{Tanabashi:2018oca}
{\bfseries Particle Data Group} Collaboration, {\em {Review of Particle
  Physics},}
\href{https://dx.doi.org/10.1103/PhysRevD.98.030001}{Phys.\  Rev.\  {\bfseries
  D98} (2018) 030001 and 2019 update}.

\bibitem{Janot:2019oyi}
P.~Janot and S.~Jadach, {\em {Improved Bhabha cross section at LEP and the
  number of light neutrino species},}
  \href{https://dx.doi.org/10.1016/j.physletb.2020.135319}{Phys.\  Lett.\
  {\bfseries B803} (2020) 135319}
{\ttfamily [\href{https://arxiv.org/abs/1912.02067}{arXiv:1912.02067}]}.

\bibitem{Schael:2013ita}
{\bfseries ALEPH, DELPHI, L3, OPAL, LEP Electroweak} Collaboration, {\em
  {Electroweak Measurements in Electron-Positron Collisions at W-Boson-Pair
  Energies at LEP},}
  \href{https://dx.doi.org/10.1016/j.physrep.2013.07.004}{Phys.\ \ Rept.\
  {\bfseries 532} (2013) 119--244} {\ttfamily
  [\href{https://arxiv.org/abs/1302.3415}{arXiv:1302.3415}]}.

\bibitem{ALEPH:2005ab}
{\bfseries ALEPH, DELPHI, L3, OPAL, SLD, LEP Electroweak Working Group, SLD
  Electroweak Group, SLD Heavy Flavour Group} Collaboration, {\em {Precision
  electroweak measurements on the $Z$ resonance},}
  \href{https://dx.doi.org/10.1016/j.physrep.2005.12.006}{Phys.\ \ Rept.\
  {\bfseries 427} (2006) 257--454} {\ttfamily
  [\href{https://arxiv.org/abs/hep-ex/0509008}{hep-ex/0509008}]}.

\bibitem{Ciuchini:2013pca}
M.~Ciuchini, E.~Franco, S.~Mishima, and L.~Silvestrini, {\em {Electroweak
  Precision Observables, New Physics and the Nature of a 126 GeV Higgs Boson},}
  \href{https://dx.doi.org/10.1007/JHEP08(2013)106}{JHEP {\bfseries 08} (2013)
  106} {\ttfamily [\href{https://arxiv.org/abs/1306.4644}{arXiv:1306.4644}]}.

\bibitem{Voutsinas:2019hwu}
G.~Voutsinas, E.~Perez, M.~Dam, and P.~Janot, {\em {Beam-beam effects on the
  luminosity measurement at LEP and the number of light neutrino species},}
  \href{https://dx.doi.org/10.1016/j.physletb.2019.135068}{Phys.\  Lett.\
  {\bfseries B800} (2020) 135068}
{\ttfamily [\href{https://arxiv.org/abs/1908.01704}{arXiv:1908.01704}]}.

\bibitem{deBlas:2019okz}
J.~de~Blas {\em et~al.}, {\em {HEPfit: a Code for the Combination of Indirect
  and Direct Constraints on High Energy Physics Models}.}
{\ttfamily \href{https://arxiv.org/abs/1910.14012}{arXiv:1910.14012}}.

\bibitem{Caldwell:2008fw}
A.~Caldwell, D.~Kollar, and K.~Kroninger, {\em {BAT: The Bayesian Analysis
  Toolkit},} \href{https://dx.doi.org/10.1016/j.cpc.2009.06.026}{Comput.\
  Phys.\  Commun.\  {\bfseries 180} (2009) 2197--2209}
{\ttfamily [\href{https://arxiv.org/abs/0808.2552}{arXiv:0808.2552}]}.

\bibitem{Awramik:2003rn}
M.~Awramik, M.~Czakon, A.~Freitas, and G.~Weiglein, {\em {Precise prediction
  for the W boson mass in the standard model},}
  \href{https://dx.doi.org/10.1103/PhysRevD.69.053006}{Phys.\ \ Rev.\ \ D
  {\bfseries 69} (2004) 053006} {\ttfamily
  [\href{https://arxiv.org/abs/hep-ph/0311148}{hep-ph/0311148}]}.

\bibitem{Awramik:2006uz}
M.~Awramik, M.~Czakon, and A.~Freitas, {\em {Electroweak two-loop corrections
  to the effective weak mixing angle},}
  \href{https://dx.doi.org/10.1088/1126-6708/2006/11/048}{JHEP {\bfseries 11}
  (2006) 048} {\ttfamily
  [\href{https://arxiv.org/abs/hep-ph/0608099}{hep-ph/0608099}]}.

\bibitem{Dubovyk:2019szj}
I.~Dubovyk, A.~Freitas, J.~Gluza, T.~Riemann, and J.~Usovitsch, {\em
  {Electroweak pseudo-observables and Z-boson form factors at two-loop
  accuracy},} \href{https://dx.doi.org/10.1007/JHEP08(2019)113}{JHEP {\bfseries
  08} (2019) 113} {\ttfamily
  [\href{https://arxiv.org/abs/1906.08815}{arXiv:1906.08815}]}.

\bibitem{Bardin:1986fi}
D.~Y.~Bardin, S.~Riemann, and T.~Riemann, {\em {Electroweak One Loop
  Corrections to the Decay of the Charged Vector Boson},}
  \href{https://dx.doi.org/10.1007/BF01441360}{Z.\  Phys.\  C {\bfseries 32}
  (1986) 121--125}.

\bibitem{Denner:1990tx}
A.~Denner and T.~Sack, {\em {The $W$-boson width},}
  \href{https://dx.doi.org/10.1007/BF01560267}{Z.\  Phys.\  C {\bfseries 46}
  (1990) 653--663}.

\bibitem{ATLAS:2018kbw}
{\bfseries ATLAS} Collaboration, {\em {A search for the rare decay of the
  Standard Model Higgs boson to dimuons in $pp$ collisions at $\sqrt{s} = 13$
  TeV with the ATLAS Detector},}
  \href{http://cds.cern.ch/record/2628763}{ATLAS--CONF--2018--026}, CERN, 2018.

\bibitem{Sirunyan:2018hbu}
{\bfseries CMS} Collaboration, {\em {Search for the Higgs boson decaying to two
  muons in proton-proton collisions at $\sqrt{s} =$ 13 TeV},}
  \href{https://dx.doi.org/10.1103/PhysRevLett.122.021801}{Phys.\  Rev.\
  Lett.\  {\bfseries 122} (2019) 021801}
{\ttfamily [\href{https://arxiv.org/abs/1807.06325}{arXiv:1807.06325}]}.

\bibitem{Seng:2018qru}
C.~Y.~Seng, M.~Gorchtein, and M.~J.~Ramsey-Musolf, {\em {Dispersive evaluation
  of the inner radiative correction in neutron and nuclear $\beta$ decay},}
  \href{https://dx.doi.org/10.1103/PhysRevD.100.013001}{Phys.\ \ Rev.\ \ D
  {\bfseries 100} (2019) 013001} {\ttfamily
  [\href{https://arxiv.org/abs/1812.03352}{arXiv:1812.03352}]}.

\bibitem{Gorchtein:2018fxl}
M.~Gorchtein, {\em {$\gamma$W Box Inside Out: Nuclear Polarizabilities Distort
  the Beta Decay Spectrum},}
  \href{https://dx.doi.org/10.1103/PhysRevLett.123.042503}{Phys.\  Rev.\
  Lett.\  {\bfseries 123} (2019) 042503}
{\ttfamily [\href{https://arxiv.org/abs/1812.04229}{arXiv:1812.04229}]}.

\bibitem{Moulson:2017ive}
M.~Moulson, {\em {Experimental determination of $V_{us}$ from kaon decays},}
  \href{https://dx.doi.org/10.22323/1.291.0033}{PoS {\bfseries CKM2016} (2017)
  033}
{\ttfamily [\href{https://arxiv.org/abs/1704.04104}{arXiv:1704.04104}]}.

\bibitem{Carrasco:2016kpy}
N.~Carrasco, {\em et al.}, {\em {$K \to \pi$ semileptonic form factors with
  $N_f=2+1+1$ twisted mass fermions},}
  \href{https://dx.doi.org/10.1103/PhysRevD.93.114512}{Phys.\ \ Rev.\ \ D
  {\bfseries 93} (2016) 114512} {\ttfamily
  [\href{https://arxiv.org/abs/1602.04113}{arXiv:1602.04113}]}.

\bibitem{Bazavov:2018kjg}
{\bfseries Fermilab Lattice, MILC} Collaboration, {\em {$|V_{us}|$ from
  $K_{\ell 3}$ decay and four-flavor lattice QCD},}
  \href{https://dx.doi.org/10.1103/PhysRevD.99.114509}{Phys.\ \ Rev.\ \ D
  {\bfseries 99} (2019) 114509} {\ttfamily
  [\href{https://arxiv.org/abs/1809.02827}{arXiv:1809.02827}]}.

\bibitem{Aguilar-Arevalo:2015cdf}
{\bfseries PiENu} Collaboration, {\em {Improved Measurement of the $\pi \to
  \mbox{e} \nu$ Branching Ratio},}
  \href{https://dx.doi.org/10.1103/PhysRevLett.115.071801}{Phys.\ \ Rev.\ \
  Lett.\  {\bfseries 115} (2015) 071801} {\ttfamily
  [\href{https://arxiv.org/abs/1506.05845}{arXiv:1506.05845}]}.

\bibitem{Czapek:1993kc}
G.~Czapek {\em et~al.}, {\em {Branching ratio for the rare pion decay into
  positron and neutrino},}
  \href{https://dx.doi.org/10.1103/PhysRevLett.70.17}{Phys.\ \ Rev.\ \ Lett.\
  {\bfseries 70} (1993) 17--20}.

\bibitem{Britton:1992pg}
D.~I.~Britton {\em et~al.}, {\em {Measurement of the $\pi^+ \to e^+$ neutrino
  branching ratio},}
  \href{https://dx.doi.org/10.1103/PhysRevLett.68.3000}{Phys.\ \ Rev.\ \ Lett.\
   {\bfseries 68} (1992) 3000--3003}.

\bibitem{Amhis:2019ckw}
{\bfseries HFLAV} Collaboration, {\em {Averages of $b$-hadron, $c$-hadron, and
  $\tau$-lepton properties as of 2018}.} {\ttfamily
  \href{https://arxiv.org/abs/1909.12524}{arXiv:1909.12524}}.

\bibitem{deBlas:2019rxi}
J.~de~Blas {\em et~al.}, {\em {Higgs Boson Studies at Future Particle
  Colliders},} \href{https://dx.doi.org/10.1007/JHEP01(2020)139}{JHEP
  {\bfseries 01} (2020) 139} {\ttfamily
  [\href{https://arxiv.org/abs/1905.03764}{arXiv:1905.03764}]}.

\bibitem{Grange:2015fou}
{\bfseries Muon g-2} Collaboration, {\em {Muon (g-2) Technical Design Report}.}
{\ttfamily \href{https://arxiv.org/abs/1501.06858}{arXiv:1501.06858}}.

\bibitem{Keshavarzi:2019bjn}
{\bfseries Muon g-2} Collaboration, {\em {The Muon $g-2$ Experiment at
  Fermilab},} \href{https://dx.doi.org/10.1051/epjconf/201921205003}{EPJ Web
  Conf.\  {\bfseries 212} (2019) 05003}
{\ttfamily [\href{https://arxiv.org/abs/1905.00497}{arXiv:1905.00497}]}.

\bibitem{Mibe:2011zz}
{\bfseries J-PARC g-2} Collaboration, {\em {Measurement of muon g-2 and EDM
  with an ultra-cold muon beam at J-PARC},}
  \href{https://dx.doi.org/10.1016/j.nuclphysbps.2011.06.039}{Nucl.\ \ Phys.\ \
  B Proc.\ \ Suppl.\  {\bfseries 218} (2011) 242--246}.

\bibitem{Abe:2019thb}
M.~Abe {\em et~al.}, {\em {A New Approach for Measuring the Muon Anomalous
  Magnetic Moment and Electric Dipole Moment},}
  \href{https://dx.doi.org/10.1093/ptep/ptz030}{PTEP {\bfseries 2019} (2019)
  053C02}
{\ttfamily [\href{https://arxiv.org/abs/1901.03047}{arXiv:1901.03047}]}.

\bibitem{Sirunyan:2018koj}
{\bfseries CMS} Collaboration, {\em {Combined measurements of Higgs boson
  couplings in proton–proton collisions at $\sqrt{s}=13\,\text {TeV}$},}
  \href{https://dx.doi.org/10.1140/epjc/s10052-019-6909-y}{Eur.\  Phys.\  J.\
  {\bfseries C79} (2019) 421}
{\ttfamily [\href{https://arxiv.org/abs/1809.10733}{arXiv:1809.10733}]}.

\bibitem{Aad:2019mbh}
{\bfseries ATLAS} Collaboration, {\em {Combined measurements of Higgs boson
  production and decay using up to $80$ fb$^{-1}$ of proton-proton collision
  data at $\sqrt{s}=$ 13 TeV collected with the ATLAS experiment},}
  \href{https://dx.doi.org/10.1103/PhysRevD.101.012002}{Phys.\  Rev.\
  {\bfseries D101} (2020) 012002}
{\ttfamily [\href{https://arxiv.org/abs/1909.02845}{arXiv:1909.02845}]}.

\bibitem{Sirunyan:2019ofn}
{\bfseries CMS} Collaboration, {\em {Search for vector-like leptons in
  multilepton final states in proton-proton collisions at $\sqrt{s}$ = 13
  TeV},} \href{https://dx.doi.org/10.1103/PhysRevD.100.052003}{Phys.\  Rev.\
  {\bfseries D100} (2019) 052003}
{\ttfamily [\href{https://arxiv.org/abs/1905.10853}{arXiv:1905.10853}]}.

\bibitem{Aad:2015dha}
{\bfseries ATLAS} Collaboration, {\em {Search for heavy lepton resonances
  decaying to a $Z$ boson and a lepton in $pp$ collisions at $\sqrt{s}=8$ TeV
  with the ATLAS detector},}
  \href{https://dx.doi.org/10.1007/JHEP09(2015)108}{JHEP {\bfseries 09} (2015)
  108}
{\ttfamily [\href{https://arxiv.org/abs/1506.01291}{arXiv:1506.01291}]}.

\bibitem{Falkowski:2013jya}
A.~Falkowski, D.~M.~Straub, and A.~Vicente, {\em {Vector-like leptons: Higgs
  decays and collider phenomenology},}
  \href{https://dx.doi.org/10.1007/JHEP05(2014)092}{JHEP {\bfseries 05} (2014)
  092}
{\ttfamily [\href{https://arxiv.org/abs/1312.5329}{arXiv:1312.5329}]}.

\bibitem{Dermisek:2014qca}
R.~Dermisek, J.~P.~Hall, E.~Lunghi, and S.~Shin, {\em {Limits on Vectorlike
  Leptons from Searches for Anomalous Production of Multi-Lepton Events},}
  \href{https://dx.doi.org/10.1007/JHEP12(2014)013}{JHEP {\bfseries 12} (2014)
  013}
{\ttfamily [\href{https://arxiv.org/abs/1408.3123}{arXiv:1408.3123}]}.

\bibitem{Ellis:2014dza}
S.~A.~R.~Ellis, R.~M.~Godbole, S.~Gopalakrishna, and J.~D.~Wells, {\em {Survey
  of vector-like fermion extensions of the Standard Model and their
  phenomenological implications},}
  \href{https://dx.doi.org/10.1007/JHEP09(2014)130}{JHEP {\bfseries 09} (2014)
  130}
{\ttfamily [\href{https://arxiv.org/abs/1404.4398}{arXiv:1404.4398}]}.

\bibitem{Kumar:2015tna}
N.~Kumar and S.~P.~Martin, {\em {Vectorlike Leptons at the Large Hadron
  Collider},} \href{https://dx.doi.org/10.1103/PhysRevD.92.115018}{Phys.\
  Rev.\  {\bfseries D92} (2015) 115018}
{\ttfamily [\href{https://arxiv.org/abs/1510.03456}{arXiv:1510.03456}]}.

\bibitem{Bhattiprolu:2019vdu}
P.~N.~Bhattiprolu and S.~P.~Martin, {\em {Prospects for vectorlike leptons at
  future proton-proton colliders},}
  \href{https://dx.doi.org/10.1103/PhysRevD.100.015033}{Phys.\  Rev.\
  {\bfseries D100} (2019) 015033}
{\ttfamily [\href{https://arxiv.org/abs/1905.00498}{arXiv:1905.00498}]}.

\bibitem{Das:2020gnt}
A.~Das, S.~Mandal, and T.~Modak, {\em {Testing triplet fermions at the
  electron-positron and electron-proton colliders using fat jet signatures}.}
  {\ttfamily \href{https://arxiv.org/abs/2005.02267}{arXiv:2005.02267}}.

\end{thebibliography}\endgroup
\end{document}